\documentclass[11pt]{article}

\usepackage{amsthm,amssymb,amsmath}

\makeatletter \@addtoreset{equation}{section} \makeatother

\newtheorem{theorem}{Theorem}
\newtheorem{lemma}{Lemma}
\newtheorem{corollary}{Corollary}
\newtheorem{remark}{Remark}

\newtheorem{proposition}{Proposition}

\def\J{{\cal J}}
\def\P{{\cal P}}

\def\ds{\displaystyle}
\def\e{\varepsilon}

\def\ti{\tilde}
\def\intd{\displaystyle\int}
\def\fracd{\displaystyle\frac}
\def\sumd{\displaystyle\sum}

\begin{document}

\title{Double scaling limit for matrix models with non analytic potentials}
\author{ M. Shcherbina\\
 Institute for Low Temperature Physics, Kharkov, Ukraine}
\date{}
\maketitle

\begin{abstract}
We study the double scaling limit for  unitary invariant
ensembles of random matrices with non analytic potentials and find the asymptotic expansion
for the entries of the corresponding Jacobi matrix. Our approach is based on
 the perturbation expansion for the string equations.
The first order perturbation terms of the Jacobi matrix coefficients are expressed
through the Hastings-McLeod solution of the Painleve II equation.
The limiting reproducing kernel is expressed in terms
of solutions of the Dirac system of differential equations with a potential
defined by the first order terms of the expansion.
\end{abstract}

\bigskip

\section{Introduction}\label{sec:1}

Unitary invariant ensembles of random matrices or
matrix models play a very important role in the random matrix theory (RMT)
mainly because of its numerous links with another fields of mathematics and theoretical
physics. An important advantage of these ensembles is that their special structure allows to study
 their limiting eigenvalue
distribution with much more details, than  other models of RMT.

 The matrix model is defined as a set of all $n\times n$ Hermitian
matrices $M$ with a probability distribution
\begin{equation}
P_{n}(M)dM=Z_{n}^{-1}\exp \{-n\mathrm{Tr}V(M)\}dM,  \label{p(M)}
\end{equation}
where $Z_{n}$ is a normalizing constant, $V:\mathbb{R}\rightarrow \mathbb{R%
}_{+}$ is a H\"{o}lder function satisfying the condition
\begin{equation}
V(\lambda )\ge (2+\epsilon )\log(1+ |\lambda |).
\end{equation}

One of important objects of the investigation in the global regime
is the Normalized Counting Measure (NCM) of eigenvalues $\{\lambda _{j}^{(n)}\}_{j=1}^n$
 of the matrix $M$. According to
\cite{BPS,Jo:98} the NCM tends weakly in probability, as $n\rightarrow
\infty $, to the non random limiting measure ${\cal N}$ known as the
Integrated Density of States (IDS). The IDS is
normalized to unity and it is absolutely continuous if $V'$ satisfies
the Lipshitz condition. The non-negative density $\rho(\lambda) $  is called  the
 Density of States (DOS) of the ensemble. The IDS can be found as a
unique solution of a certain variational problem
\cite{BPS,Jo:98} which imply, in particularly,
 that  if $V'(\lambda)$  satisfies
the Lipshitz conditions on the support $\sigma$ of limiting IDS, then
$\rho(\lambda)$ is a solution of the following integral equation
\begin{equation} \label{rep-V}
V'(\lambda)=2\int_{\sigma}\frac{\rho(\mu)d\mu}{\lambda-\mu},\quad \sigma=\hbox{supp}{\cal N}.
\end{equation}

While the global regime  depends strongly on the form of $V$, the local eigenvalue
statistics  is expected to be universal.
Denote by $p_{n}(\lambda _{1},...,\lambda _{n})$ the joint
eigenvalue probability density. It is known (see \cite{Me:91}) that
\begin{equation}
p_{n}(\lambda _{1},...\lambda _{n})=Z_{n}^{-1}\prod_{1\leq j<k\leq
n}(\lambda _{j}-\lambda _{k})^{2}\prod_{j=1}^{n}e^{-nV(\lambda
_{j})}, \label{p(la)}
\end{equation}
where $Z_{n}$ is the respective normalization factor. Let
\begin{equation}
p_{l}^{(n)}(\lambda _{1},...,\lambda _{l})=\int p_{n}(\lambda
_{1},...,\lambda _{l},\lambda _{l+1},...\lambda _{n})d\lambda
_{l+1}...d\lambda _{n}  \label{dp_k}
\end{equation}
be the $l$th marginal distribution density of (\ref{p(la)}).
Universality of local eigenvalue statistics means that
if we consider some $\lambda_0\in\sigma$, then all marginal distribution
densities after a proper rescaling  (which depends on the behavior of the limiting DOS
$\rho(\lambda)$ near the point $\lambda=\lambda_0$) tend to some universal limits.

The most known quantity probing  universality is the large-$n$ form of
the hole probability
\begin{equation}
E_{n}(\Delta _{n})=\mathbf{P}_{n}\{\lambda _{l}^{(n)}\notin \Delta
_{n},\;l=1,...,n\},  \label{hon}
\end{equation}%
where $\mathbf{P}_{n}\{...\}$ is the probability defined by the distribution
(\ref{p(M)}), and $\Delta _{n}$ is an interval of the spectral
axis, whose order of magnitude is fixed by the condition \quad $n{\cal N}(\Delta
_{n})\sim 1$. For the matrix models $E_{n}(\Delta _{n})$ can be obtained as the
Fredholm determinant of a certain integral operator.
 This structure of the hole probability  is a consequence
  of the structure of marginal densities and the latter can be explained
  by  the link
 of matrix models with orthogonal polynomials
 $p_{l}^{(n)}(\lambda ),\;(l=1,...)$  on $\mathbb{R}$ associated with the weight
 $e^{-nV(\lambda )}$.
The link is provided by the formula
\cite{Me:91}
\begin{equation}
p_{l}^{(n)}(\lambda _{1},...,\lambda _{l})={\frac{(n-l)!}{n!}}\det
||K_{n}(\lambda _{j},\lambda _{k})||_{j,k=1}^{l},  \label{p_k=}
\end{equation}
where
\begin{equation}
K_{n}(\lambda ,\mu )=\sum_{l=1}^{n}\psi _{l}^{(n)}(\lambda )\psi
_{l}^{(n)}(\mu )  \label{k_n}
\end{equation}
is known as a reproducing kernel of an orthonormalized system
\begin{equation}
\psi _{l}^{(n)}(\lambda )=e^{-nV(\lambda )/2}P_{l-1}^{(n)}(\lambda
),\;\,l\in\mathbb{N},  \label{psi}
\end{equation}
in which $P_{l}^{(n)}(\lambda )$ is a polynomial of $l$-th degree
with a positive coefficient in front of $\lambda ^{l}$. This
polynomial is uniquely  defined by
 the orthogonality conditions
\begin{equation}
\int P_{l}^{(n)}(\lambda )P_{m}^{(n)}(\lambda
)e^{-nV(\lambda)}d\lambda =\delta _{l,m}.  \label{ortP}
\end{equation}

Formula (\ref{p_k=}) gives us
\[
(n^{1-\gamma}/c_0)^{\ell}p_{\ell,n}(\lambda_0+t_1/c_0n^{\gamma},\dots,
\lambda_0+t_{\ell}/c_0n^{\gamma})=\hbox{det}|{\cal
K}_n(t_i,t_j)|_{i,j=1}^\ell,\] where
\[{\cal K}_n(t_1,t_2)=n^{1-\gamma}c_0^{-1}
K_n(\lambda_0+t_1 (c_0n)^{-\gamma}, \lambda_0+t_2
(c_0n)^{-\gamma}).
\]
Hence we can   reduce the question on the behaviour
of the rescaled $\ell$th marginal density to the question  of the existence of the limit of
${\cal K}_n(s,t)$ for  proper chosen $\gamma$ and $c_0$.

In the bulk case ($\rho(\lambda_0)\not=0$) we choose $\gamma=1$.
Then the limiting hole probability is the Fredholm determinant of
the integral operator, defined by the kernel $\sin \pi
(t_{1}-t_{2})/\pi (t_{1}-t_{2})$ on the interval $(0,s)$. This
fact for the GUE was established by M. Gaudin in the early 60s
\cite{Me:91}. The same fact was proved recently in
\cite{PS:97,De-Co:99} for certain classes of matrix models. In
terms of the reproducing kernel (\ref{k_n}) this result can be
formulated  as
\[
\lim_{n\to\infty}\rho^{-1}(\lambda_0)
K_n(\lambda_0+t_1/\rho(\lambda_0)n,\lambda_0+t_2/\rho(\lambda_0)n)
=\frac{\sin\pi(t_1-t_2)}{\pi(t_1-t_2)}.
\]
The edge case of local eigenvalue statistics was studied much later even for
the GUE \cite{Fo:93,Tr-Wi:94a}. It was found that if we choose
$\gamma=2/3$, then
for the  edge points $\lambda_0=\pm a$ ( $\sigma=[-a,a]$) the hole probability (\ref{hon}) of the
GUE in the limit $n\rightarrow \infty $ is the Fredholm determinant of the integral operator,
 defined on the
interval $(0,s)$ by the kernel
\[
\mathcal{K}(t_1,t_2)=\frac{Ai(t_1)Ai^{\prime }(t_2)-Ai^{\prime
}(t_1)Ai(t_2)}{t_1-t_2},
\]
where $Ai(s)$ is a standard Airy function \cite{AS}.
This fact for real analytic potentials in (\ref{p(M)}) was obtained in
 \cite{De-Co:99a}.  In the paper \cite{PS3} a more simple proof of the edge
universality  for the same class of
potentials was given. An important
advantage of the method  of \cite{PS3} is that it can be generalized
to a class  of non analytic potentials.

\smallskip
The case of the critical point universality with
$\gamma=1/3$ was studied first for
$V(\lambda)=\frac{1}{4}\lambda^4-\lambda^2$ by using the Riemann-Hilbert
approach in \cite{BI}.
 The same method was generalized on a class of real analytic potentials in
 \cite{CK} under  additional assumptions that the limiting spectrum $\sigma$ consists
 of one interval and the density $\rho(\lambda)$ behaves like a square root near the edge
 points and has only one critical point inside $\sigma$ (cf. condition {\bf C3} below).
But the asymptotic behaviour of the Jacobi matrix coefficients was not studied.

In the present paper we find the asymptotic behaviour of the Jacobi matrix coefficients
and on the basis of this result prove  universality  near
the critical  point.
We need not to assume that $V(\lambda)$ is a real analytic function.
 Our approach is based on the mathematical version of physical ideas proposed in
 \cite{Bo-Br:91}.

Let us state  our main conditions.
\begin{description}
 \item[\textbf{ C1.}]\textit{  The support $\sigma $ of the
IDS of the ensemble consists of a single interval:}
\[
\sigma =[-2,2].
\]
\item[\textbf{ C2.} ] \textit{$V(\lambda )$ is an even real locally Lipshitz function
    in $ \mathbb{R}$ and  there exists $\e>0$  such that
    $V^{(6)}(\lambda )\in L_2[\sigma_\epsilon]$, where $\sigma_\e=[-2-\e,2+\e]$.}
\item[\textbf{C3.}]
\textit{The DOS $\rho(\lambda)$ has the form}
\begin{equation}\label{P1}
\rho (\lambda )=\frac{1}{2\pi}\lambda^2
P_0(\lambda )\sqrt{4-\lambda ^{2}},\quad \lambda\in [-2,2],
\end{equation}
\textit{where $P_0(\lambda)>0$ for  $\lambda\in[-2,2]$.}

\item[\textbf{C4.}]\textit{ The function
\begin{equation}
u(\lambda)=2\int\log |\mu-\lambda|\rho(\mu)d\mu-V(\lambda)
\label{u}\end{equation}
 achieves its maximum
if and only if $\lambda\in[-2,2]$.}
\end{description}

\begin{remark}
\label{rem:1} In fact Theorems 1 and 2 below can be proved under
the condition $V^{(5)}(\lambda )\in L_2[\sigma_\epsilon]$, but the
proof is more complicated. Since $V^{(4)}(0)$ is used in the
limiting formulas for Theorem \ref{thm:1}, it is natural to expect
that existence of continuous $V^{(4)}(\lambda)$ in some
neighborhood of $\lambda=0$ is a necessary condition for Theorem
\ref{thm:1}. Thus condition $\mathbf{C2}$ does not look too
restrictive.
\end{remark}
\begin{remark}\label{rem:3}
It is well known DOS $\rho $ of the ensemble
(\ref{p(M)}) for $\sigma =[-2,2]$  has the form
\begin{equation}
\rho (\lambda )=\frac{1}{2\pi}\chi _{\sigma }(\lambda ) P(\lambda
)\sqrt{4-\lambda ^{2}},  \label{rho}
\end{equation}
where $\chi _{\sigma }(\lambda )$ is the indicator of
$\sigma $
and it follows from (\ref{rep-V}) that $P(\lambda)$  can be represented in the form
\begin{equation}
P(\lambda)={\frac{1}{\pi }}\int_{\sigma}\frac{V'(\lambda)-
V'(\mu)}{(\lambda-\mu)\sqrt{4-\mu ^{2}}}d\mu , \label{P}
\end{equation}
So condition $\mathbf{C3}$ means that $\rho(\lambda)$,
behaves like square root near the edge points and has the second order zero
at $\lambda=0$.
\end{remark}
Define a semi infinite Jacobi matrix $\mathcal{J}^{(n)}$,
 whose entries $J_{l+1,l}^{(n)}=
J_{l,l+1}^{(n)}=J_{l}^{(n)}$ are defined by the recurrent relations
 \begin{equation}
J_{l}^{(n)}\psi _{l+1}^{(n)}(\lambda )+J_{l-1}^{(n)}\psi _{l-1}^{(n)}(\lambda )
=\lambda \psi _{l}^{(n)}(\lambda
),\quad l\in\mathbb{N}, \quad J_{0}^{(n)}=0,\label{rec}
\end{equation}
where $\psi _{l}^{(n)}$ is defined by (\ref{psi}).
The main result of the paper is
\begin{theorem}\label{thm:1}
Let conditions C1--C4 be fulfilled. Then for any $k:|k|\le n^{1/3}\log^2 n$
\begin{equation}
J_{n+k}^{(n)}=1+\tilde s(-1)^kn^{-1/3}q(\frac{k}{n^{1/3}})
+\frac{k}{8nP_0(2)}+ r_k, \label{t1.1}\end{equation}
where $q(x)$
 is the Hastings-McLeod solution of the Painleve II equation
\begin{equation}
q''(x)=\frac{1}{2P_0(0)}xq(x)+2q^3(x), \label{Pain}\end{equation}
which is uniquely defined (see \cite{HMcL}) by the asymptotic conditions
\begin{equation}
\lim_{x\to +\infty}q(x)=0,\quad \lim_{x\to
-\infty}\frac{q(x)}{(-x)^{1/2}}=\frac{1}{2P^{1/2}_0(0)},
\label{as-Pain}\end{equation}
 $P_0(\lambda)$ is defined by (\ref{P1}), $\tilde s=\emph{sign}(1-J_{n}^{(n)})$
 and  remainder terms $r_k$ satisfy
the bounds
\begin{equation}\label{r_k}
|r_k|\le C\bigg(\left|k/n\right|^{2}+n^{-4/3}\bigg),
\end{equation}
where $C$ is some absolute constant.
\end{theorem}
To prove universality of local eigenvalue statistics we study
\begin{equation}\label{def-K}
  {\cal K}_n(t_1,t_2)=n^{-1/3}K_n(t_1 n^{-1/3}, t_2n^{-1/3}).
\end{equation}

\begin{theorem}\label{thm:2}
Under conditions C1--C4  for any $l\in\mathbb{N}$ there exists a weak limit
of the marginal density (\ref{dp_k})
\begin{equation}\label{lim-K}
 \lim_{n\to\infty}(2n^{2/3})^lp^{(n)}_l(2t_1/n^{2/3},\dots,2t_l/n^{2/3})=
 \emph{det}\{\mathcal{K}(t_i,t_j)\}_{i,j=1}^l,
\end{equation}
where
\begin{equation}\label{exp_K}\mathcal{K}(t_1,t_2)=\frac{\Psi_1(0;t_1)\Psi_0(0;t_2)-
\Psi_0(0;t_1)\Psi_1(0;t_2)}{\pi(t_1-t_2)},
 \end{equation}
and $\mathbf{\Psi}(x,t)=(\Psi_0(x;t),\Psi_1(x;t))$ is a solution of the Dirac system of
equations
\begin{equation}\label{A}
{\cal A}\mathbf{\Psi}(x,t)=t\mathbf{\Psi}(x,t),\quad
 {\cal A}=\left(\begin{array}{cc}
   0 & 1 \\
   -1 & 0\
 \end{array}\right)\frac{d}{dx}+\left(\begin{array}{cc}
   0 & q(x) \\
   q(x) & 0\
 \end{array}\right),
\end{equation}
with $q(x)$ defined by (\ref{Pain})-(\ref{as-Pain}), and
$\mathbf{\Psi}(x,t)$ are chosen to satisfy the asymptotic conditions
\begin{equation}\label{as-Psi}
  \lim_{x\to -\infty}|\mathbf{\Psi}(x;t)|=0,\quad
\lim_{x\to\infty}|\mathbf{\Psi}(x;t)|=1.
\end{equation}
\end{theorem}
\begin{corollary}\label{cor:1}
Under conditions C1--C4  the hole probability
(\ref{hon}) for $\Delta_n=[n^{-1/3}a,n^{-1/3}b]$ converges, as $n\to\infty$,
to the Fredholm determinant  of the integral operator defined in $[a,b]$ by the kernel
(\ref{lim-K}):
\begin{equation}\label{hole}
\lim_{n\to\infty}E_n([an^{-1/3},bn^{-1/3}])=\emph{det}(I-\mathcal{K}([a,b])).
\end{equation}
\end{corollary}

The paper is organized as follows. In Section \ref{sec:2} we prove
Theorems \ref{thm:1} and \ref{thm:2}. The proofs of the most of auxiliary
results are given
in Section \ref{sec:3}. Some auxiliary results which have no direct
links with  matrix models (some properties of the Hastings-McLeod
solution, bounds for smooth functions of Jacobi matrices etc.),
are proven in Appendix.

\section{Proofs of Theorems \ref{thm:1}, \ref{thm:2}}\label{sec:2}

{\it Proof of Theorem \ref{thm:1}.}
The main idea of the proof is to use the perturbation expansion of the
string equations:
\begin{equation}
J_{k}^{(n)}V^{\prime }({\cal J}^{(n)})_{k,k+1} =\frac{k}{n},
\label{string}
\end{equation}
which we consider  as a system of nonlinear equations
with respect to the coefficients $J_{k}^{(n)}$.
Here and below  we denote by $\J^{(n)}$  a semi-infinite Jacobi matrix,
 defined in (\ref{rec}).
Relations (\ref{string})  can be easily obtained from the identity
\[
\int \left( e^{-nV(\lambda )}P_{k+1}^{(n)}(\lambda
)P_{k}^{(n)}(\lambda )\right) ^{\prime }d\lambda =0.
\]
 Our first step is the following lemma, proven in Section \ref{sec:3}:

\begin{lemma}\label{lem:1}
Under conditions $\mathbf{C1- C3}$ uniformly in $k:|k-n|\le n^{1/2}$
\[  |J^{(n)}_{k}-1|\le Cn^{-1/8}\log^{1/4} n,\quad
 |J^{(n)}_{k}+J^{(n)}_{k+1}-2|\le Cn^{-1/4}\log^{1/2} n.
\]
\begin{remark} \label{rem:4}
The convergence $J^{(n)}_{k}\to 1$, as $n\to\infty$ and $|k-n|=o(n)$
 without  uniform bounds for the remainders was proven in \cite{APS} under much more
weak conditions ($V'(\lambda)$ is a H\"{o}lder function in some neighborhood of the
limiting spectrum).
\end{remark}
\end{lemma}
Denote ${\cal J}^{(0)}$ an infinite Jacobi matrix with constant coefficients
\begin{equation}\label{J^0}
  \J^{(0)}_{k,k+1}= \J^{(0)}_{k+1,k}=1
\end{equation}
and for any positive $N<n$ define   an infinite Jacobi matrix
$\tilde \J(N)$ with the entries
\begin{equation}\label{ti-J}
 \ti J_k=
  \begin{cases}
   J^{(n)}_{n+k}-1,  & |k|< N , \\
    0, & \text{otherwise}.
  \end{cases}
\end{equation}
\begin{proposition}\label{pro:2}
For any  function $ v(\lambda)$,
whose $\ell$th  derivative belongs  to $L_2[\sigma_\e]$ ($\sigma_\e=[-2-\e,2+\e]$),
 consider a periodic function $\tilde v(\lambda)=\tilde v(\lambda+4+2\e)$
 with the same number of derivatives, and such that
$\tilde v(\lambda)=v(\lambda)$ for $ |\lambda|\le 2+\e/2$.
Let also $n^{1/2}\ge N,M>n^{1/3}$ and $\tilde\J(N+M)$ is defined by (\ref{ti-J}). Then
uniformly in $N$, $M$ and $|k|\le N$ for any fixed integer $\delta$
\begin{equation}\label{v(J)}
 v(\J^{(n)})_{n+k,n+k+\delta}-\tilde v(\J^{(0)}+\tilde \J(N+M))_{k,k+\delta}=O(M^{-\ell+1/2}).
\end{equation}
\end{proposition}
\noindent The proof of the proposition is given in Appendix.

 To estimate the remainder terms of our expansion we define
\begin{equation}\label{m_N}
  m_k:=\max\bigg\{\max_{|j|\le |k|+n^{1/3}/2}\left\{|\ti J_{j}|,
  |\ti J_{j}+\ti J_{j+1}|^{1/2}\right\},
  \left(|k|/n\right)^{1/2}\bigg\}.
\end{equation}
\begin{lemma}\label{lem:2}
Let $v(\lambda)$ satisfy conditions of Proposition \ref{pro:2}
 with $\ell=5$,   $\delta$  be any fixed  integer and  $|k|\le 3n^{1/2}/4$. Then
\begin{equation} \begin{array}{rcl}
v(\J^{(n)})_{n+k,n+k+\delta}&=&
 v(\J^{(0)})_{k,k+\delta}-c_1^{(\delta)}\ti J_k+\sum\nolimits'
 \P_{k-l_1}^{(\delta)}\ti J_{l_1}
  +\sum\nolimits'\P^{(2,k,\delta)}_{l_1,l_2}\tilde J_{l_1}\tilde J_{l_2}\\&&
+\sum\nolimits'\P^{(3,k,\delta)}_{l_1,l_2,l_3}\tilde J_{l_1} \tilde
J_{l_2}\tilde J_{l_3}+r_k^{(\delta)}\\&=&
v(\J^{(0)})_{k,k+\delta}-c_1^{(\delta)}\ti
J_k+\Sigma^{(1)}+\Sigma^{(2)}+\Sigma^{(3)} +r_k^{(\delta)},
\end{array}\label{exp-v}\end{equation}
where  $|r_k^{(\delta)}|\le Cm^{4}_k$
and  $\P^{(2,k,\delta)}_{l_1,l_2}$ and
$\P^{(3,k,\delta)}_{l_1,l_2,l_3}$ satisfy the bounds:
\begin{equation}\label{b-P}\begin{array}{l}
\bigg|\sum'\P^{(2,k,\delta)}_{l_1,l_2}(l_1-k)(l_2-k) \tilde x_{l_1}\tilde
y_{l_2}\bigg|\le C \,||\tilde x||\,||\tilde y||,\\
\bigg|\sum'\P^{(2,k,\delta)}_{l_1,l_2}(l_1-k)^2 \tilde x_{l_1}\tilde
y_{l_2}\bigg|\le C\,||\tilde x||\,||\tilde y||,\\
\bigg|\sum'\P^{(3,k,\delta)}_{l_1,l_2,l_3}(l_1-k) \tilde x_{l_1}\tilde
y_{l_2}\tilde z_{l_3}\bigg| \le C\,||\tilde x||\,||\tilde
y||\,||\tilde z||
\end{array}\end{equation}
for any bounded sequences $\{\tilde x_k\}$, $\{\tilde y_k\}$ and $\{\tilde z_k\}$. Here
 and below $||x||=\max_{k}|x_k|$ and $\sum'$ means the summation over indexes $|l_i|\le
 |k|+n^{1/3}/2$

 Moreover,
\begin{equation}\label{P_l}
\quad \P_l^{(\delta)}=\frac{1}{2\pi}\int_{-\pi}^{\pi}
F^{(\delta)}(2\cos(x/2)) e^{ilx}
dx,\end{equation}
with some smooth $F^{(\delta)}(\lambda)$ and for $\delta=1$
\begin{equation}\label{P^1}
c_1^{(1)}=\fracd{1}{2\pi}\int_{-\pi}^{\pi} v(2\cos x)\cos x\,dx,\quad
F^{(1)}(\lambda)=2P(\lambda)=\fracd{2}{\pi}\int_{-2}^2\frac{v(\lambda)-v(\mu)}
{(\lambda-\mu)\sqrt{4-\mu^2}}d\mu.
\end{equation}
For $\delta=0$
\begin{equation}\label{F^0}
c_1^{(0)}=0,
 \quad F^{(0)}(2\cos(x/2))=\frac{\cos^2(x/2)}{2\pi}\int_{-\pi}^{\pi}
 \frac{v(2\cos x_1)dx_1}{\cos^2x_1-\cos^2(x/2)}.
\end{equation}
\end{lemma}
\noindent The proof of the lemma is given in Section \ref{sec:3}.
\begin{remark}\label{rem:5} If $v$ coinsides with $V'$
for $\lambda\in\sigma_{\e/2}$, then
\begin{equation}\label{int0}
 c_1=\frac{1}{2\pi}\int_{-\pi}^{\pi} v(2\cos x)\cos x\,dx=
\frac{1}{\pi}\int_{-\pi}^{\pi}dx\int_{-2}^{2}\cos
x\frac{\rho(\lambda)d\lambda}{2\cos x-\lambda}=1.
\end{equation}
\end{remark}
\begin{remark}\label{rem:6}
If in (\ref{P^1}) $P(\lambda)=\lambda^2P_0(\lambda)$, then for any $\ti x_k$
\begin{equation}\label{P_1}
\sum\P_{k-l}^{(1)}\ti x_l=\sum\P^{(0)}_{k-l}(\ti x_{l+1}+2\ti x_l+\ti
x_{l-1}),\end{equation}
where
\[
\P^{(0)}_{l}=\frac{1}{\pi}\int_{-\pi}^{\pi}P_0(2\cos (x/2))
e^{ilx} dx.
\]
\end{remark}
Denote
\begin{equation}\label{y_k}
  y_k=k/(8P_0(2)n),
\end{equation}
where $P_0(\lambda)$ is defined by (\ref{P1}). Then we represent $\ti
J_k$ in the form
\begin{equation}\label{x_k}
  \ti J_k=(-1)^kx_k+y_k, \quad or \quad x_k=(-1)^k(\ti J_k-y_k).
\end{equation}
Now, substituting (\ref{x_k}) in (\ref{exp-v}) and keeping the
terms up to the order $m^3_k$ (recall, that by definition (\ref{m_N})
$y_k=O(m^2_k)$), we get for $\delta=1$
\begin{equation}\label{t1.5}\begin{array}{rcl}
\Sigma^{(1)}&=&-\sum'\P^{(0)}_{k-l}(-1)^{l}
d^{(2)}_l+\sum'\P^{(1)}_{k-l}y_l=-\Sigma^{(1)}_1+\Sigma^{(1)}_2,\\
\Sigma^{(2)}&=&\sum'\P^{(2,k,1)}_{l_1,l_2}(-1)^{l_1+l_2}x_{l_1}x_{l_2}+
2\sum'\P^{(2,k,1)}_{l_1,l_2}(-1)^{l_1}x_{l_1}y_{l_2}
+ O(m^{4}_k)\\ &=&\Sigma^{(2)}_1 +\Sigma^{(2)}_2+ O(m^{4}_k),\\
\Sigma^{(3)}&=&\sum'\P^{(3,k,1)}_{l_1,l_2,l_3}(-1)^{l_1+l_2+l_3}
x_{l_1}x_{l_2}x_{l_3}+ O(m^{4}_k) =\Sigma^{(3)}_1+ O(m^{4}_k).
\end{array}\end{equation}
Here and below we denote
\begin{equation}\label{d}
  d^{(1)}_k=x_{k+1}-x_{k},\quad d^{(2)}_k=d^{(1)}_k-d^{(1)}_{k-1},
  \quad d^{(3)}_k=d^{(2)}_{k+1}-d^{(2)}_{k}.
\end{equation}
Since by  definition (\ref{m_N}) $|d^{(1)}_k|\le Cm^2_k$, using
bounds (\ref{b-P}), we can write
\begin{equation}\label{t1.6}\begin{array}{rcl}
\Sigma^{(2)}_1&=&
x_{k}^2\sum'\P^{(2,k,1)}_{l_1,l_2}(-1)^{l_1+l_2}+
2x_{k}d^{(1)}_k\sum'\P^{(2,k,1)}_{l_1,l_2}(-1)^{l_1+l_2}(l_1-k)
\\&&+2x_{k}\sum'\P^{(2,k,1)}_{l_1,l_2}(-1)^{l_1+l_2}
\bigg((x_{l_1}-x_k)-(l_1-k)d^{(1)}_{k}\bigg) + O(m^{4}_k)\\
&=&x_k^2\Sigma^{(2)}_{1,1}
+2x_kd^{(1)}_k\Sigma^{(2)}_{1,2}+2x_k\Sigma^{(2)}_{1,3}+
O(m^{4}_k).
\end{array}\end{equation}
Similarly
\begin{equation}\label{t1.7}\begin{array}{l}
\Sigma^{(2)}_2=
x_ky_k\sum'\P^{(2,k,1)}_{l_1,l_2}(-1)^{l_1}+
O(m^{4}_k) =x_ky_k\Sigma^{(2)}_{2,1}+ O(m^{4}_k)
\\
\Sigma^{(3)}_1
=x_k^3\sum'\P^{(3,k,1)}_{l_1,l_2,l_3}(-1)^{l_1+l_2+l_3}
+ O(m^{4}_k)=x_k^3\Sigma^{(3)}_{1,1}+ O(m^{4}_k)
\end{array}\end{equation}
and
\[\Sigma^{(1)}_1=-d^{(2)}_k\sum\nolimits'\P^{(0)}_{k-l}(-1)^{l}-
\sum\nolimits'\P^{(0)}_{k-l}(-1)^{l}(d^{(2)}_l-d^{(2)}_k)=-d^{(2)}_k
\Sigma^{(1)}_{1,1}-\Sigma^{(1)}_{1,2}.
\]
\begin{proposition}\label{pro:3} If $v(\lambda)=V'(\lambda)$, as
$\lambda\in\sigma_{\e/2}$, then
\begin{equation}\label{p3.1}\begin{array}{ll}
\Sigma^{(1)}_{1,1}=2(-1)^kP_0(0)+O(n^{-5/6}),
&\Sigma^{(1)}_2=k/n+O(n^{-3/2}),
\\
\Sigma^{(2)}_{1,1}=1+O(n^{-5/6}),&\Sigma^{(2)}_{1,2}=O(n^{-1/2}),\\
\Sigma^{(2)}_{2,1}=(-1)^k+O(n^{-5/6}),&
\Sigma^{(3)}_{1,1}=(-1)^k(4P_0(0)-1) +O(n^{-1/2}).
\end{array}\end{equation}
If $v^{(0)}(\lambda)=\lambda^{-1}V'(\lambda)$ for
$\lambda\in\sigma_{\e/2}$, then
\begin{equation}\label{p3.1a}
\sum\nolimits'\P^{(0)}_{l-k}=4P_0(2)+O(n^{-5/6}),\quad
\sum\nolimits'\P^{(2,k,0)}_{l_1,l_2}(-1)^{l_1+l_2}=2P_0(0)+O(n^{-5/6}).
\end{equation}
\end{proposition}
Substituting (\ref{p3.1}) into (\ref{t1.6})-(\ref{t1.7}), we
obtain
\begin{multline}\label{t1.9a}
V'(\J^{(n)})_{n+k,n+k+1}=1-(-1)^kx_k-y_k-\sum\nolimits'\P^{(0)}_{k-l}(-1)^{l}
d^{(2)}_l+k/n\\ +x_k^2+2x_k\Sigma^{(2)}_{1,3}+
2x_ky_k(-1)^k+(-1)^k(4P_0(0)-1)x_k^3+O(m^{4}_k).
 \end{multline}
Using this expression in (\ref{string}) and keeping the terms up
to
 the order $O(m^3_k)$, we get
\begin{multline}\label{t1.9}
-\sum\nolimits'\P^{(0)}_{k-l}(-1)^{(l-k)}
d^{(2)}_l+4P_0(0)x_k^3+8P_0(2)x_ky_k +2x_{k}\Sigma^{(2)}_{1,3}=
O(m^{4}_k).
\end{multline}
\begin{remark}\label{rem:7}
If the operator $\P^{(0)}$ has the form (\ref{P_1}) with  $P_0(x)>d>0$,
 then there exists
\[
(\P^{(0)})^{-1}_{k,l}=\frac{1}{4\pi}\int_{-\pi}^\pi
P_0^{-1}(2\cos(x/2))e^{i(k-l)x}dx.
\]
So, if for some $z_j$ we have the system of equations
\begin{equation}\label{l_eq}
 \sum\P^{(0)}_{k-l} z_l =\e_k,\quad k\in\mathbb{Z},
\end{equation}
then
\begin{equation}\label{b_sol}
| z_k|  \le \sum_{l}|(\P^{(0)})^{-1}_{k-l}||\e_l|.
\end{equation}
\end{remark}
We apply this remark to (\ref{t1.9}) written in the form
(\ref{l_eq}) with $z_k=(-1)^kd^{(2)}_k$, where $d^{(2)}_k$ are
defined by (\ref{d}) with $x_k$ of the form (\ref{x_k}) for
$|k|\le n^{1/2}$ and $x_k=0$ for $k>n^{1/2}$. We take also
\begin{equation}\label{e_k}
\e_k=
  \begin{cases}
   4P_0(0)x_k^3+8P_0(2)x_ky_k
+2x_{k}\Sigma^{(2)}_{1,3}+\sum_{|l|>N_k}P^{(0)}_{k-l}(-1)^{l}
d^{(2)}_l+ O(m^{4}_k),  &k\le 3n^{1/2}/4 , \\
   \sum P^{(0)}_{k-l}(-1)^{l}
d^{(2)}_l,  & \text{otherwise}.
  \end{cases}
\end{equation}
Since $P^{(0)}(\lambda)$ has the third derivative (see conditions
C2, C3 and representation (\ref{P}), $(P^{(0)}(\lambda))^{-1}$
also does. Hence, using the standard bound for the remainder of
the Fourier expansion of the functions $P^{(0)}(\lambda)$ and
$(P^{(0)}(\lambda))^{-1}$ (see, e.g. (\ref{Four1})), we have for
any $M$
\begin{equation}\label{e_k.1}\sum_{|l|>M}|\P^{(0)}_{l}|\le CM^{-5/2},\quad
\sum_{|l|>M}|(\P^{(0)})^{-1}_{l}|\le CM^{-5/2}.
\end{equation}
Using the first of these bounds and the inequality
\[
|d^{(2)}_k|=|d^{(1)}_k-d^{(1)}_k|\le\left\{\begin{array}{ll}
Cn^{-1/4}\log^{1/4}n,& ||k|- n^{1/2}|>2\\Cn^{-1/8}\log^{1/8}n,&
otherwise\end{array}\right.
\]
which follows from Lemma \ref{lem:1}, we get
\begin{equation}\label{e_k.2}
|\e_k|\le
 C \begin{cases}
  m_k^3+n^{-13/12}\log^{1/2}n,   &|k|\le n^{1/2}/2 , \\
    n^{-1/4}\log^{1/2}n &n^{1/2}/2<|k|\le n^{1/2}/2\\
   Cn^{-1/8}\log^{1/4}n, & \text{otherwise}.
  \end{cases}
\end{equation}
Using this bound in the l.h.s. of (\ref{b_sol}) and taking into
account the second bound in (\ref{e_k.1}), we obtain
\begin{equation}\label{b-d^2}
|d^{(2)}_k|\le Cm^{3}_{k+[n^{1/3}/4]},\quad k\le n^{1/2}/3.
\end{equation}
Hence, using this bound in (\ref{b_sol}) and (\ref{b-P}), we get
that $x_k\Sigma^{(2)}_{1,3}$ (see (\ref{t1.6})) is less than
$O(m^{4}_{k+[n^{1/3}/4]})$, and therefore (\ref{t1.9}) can be
rewritten as
\begin{equation}\label{t1.10}
-\sum\nolimits'\P^{(0)}_{k-l}(-1)^{l}
d^{(2)}_l+4P_0(0)x_k^3+8P_0(2)x_ky_k =O(m^{4}_{k+[n^{1/3}/4]}).
\end{equation}
Now subtracting from (\ref{t1.10}) the same equation written for
$k:=k-1$, we get
\[
\sum\nolimits'\P^{(0)}_{k-l}(-1)^{l} d^{(3)}_l
=O(m^{4}_{k+[n^{1/3}/4]}).
\]
Using Remark \ref{rem:7} by the way described above, we obtain
that  $|d^{(3)}_k|\le C m^{4}_{k+[n^{1/3}/2]}$ for $|k|\le
n^{1/2}/4$. Hence, writing
\[\sum\nolimits'\P^{(0)}_{k-l}(-1)^{l}
d^{(2)}_l=d^{(2)}_k\Sigma^{(1)}_{1,1}+\sum\nolimits'\P^{(0)}_{k-l}(-1)^{l}
(d^{(2)}_l-d^{(2)}_k)=d^{(2)}_k\Sigma^{(1)}_{1,1}+O(m^{4}_{k+[n^{1/3}/2]}),
\]
in view of the first relation in (\ref{p3.1}), we get
from (\ref{t1.10})
\begin{equation}\label{t1.11}
 d^{(2)}_k-2x_k^3-\frac{k}{2P_0(0)n}x_k= r_k, \quad
|r_k|\le \tilde C'\tilde m^{4}_k,\quad where\quad
 \tilde m_k:=m_{k+[n^{1/3}/2]}.
\end{equation}
\begin{lemma}\label{lem:4}
There exist $C^{*},L^*>0$ such that for any $k: n^{1/2}/5>|k|>L^*n^{1/3}$
\begin{equation}\label{l4.1}
\tilde m_k \le C^*\left(|k|/n\right)^{1/2}.
\end{equation}
Besides, there exist $C_{1,2,3}$ such that for
$n^{1/3}<k<k^*=[n^{1/3}\log^2 n]$
\begin{equation}\label{as+}
|x_k| \le C_1n^{-1/3}e^{-C_2(k/n^{1/3})^{3/2}}+C_3\tilde
m^{4}_{2k^*}.
\end{equation}
\end{lemma}
The proof is given in the next section. It is based
on the  proposition  proven  in Appendix.
\begin{proposition}\label{p:d^2} Let  $\{\tilde x_k\}_{|k|<M}$,
satisfy the recursive relations:
\begin{equation}\label{d^2>}
\tilde x_{k+1}-2\tilde x_{k}+\tilde x_{k-1}
=2\tilde x_{k}^3+\tilde r_k,\quad |\tilde r_k|\le \varepsilon^3,
\quad |\tilde x_{k}|\le \varepsilon_1.
\end{equation}
Then for any  $|k|<M-2M_1$ with $M_1>2\e^{-1}/3$
\begin{equation}\label{p5.0}
|\tilde x_{k}|\le \max\{\e,(2M_1^{-2}\e_1)^{1/3}\},\quad |\tilde
x_{k+1}-\tilde x_{k}|\le 4\max\{\e^2,(2M_1^{-2}\e_1)^{2/3}\}.
\end{equation}
If for $|k|\le M$
\begin{equation}\label{d^2}\tilde x_{k+1}-2\tilde x_{k}+\tilde x_{k-1}=
f_k\tilde x_{k}+\tilde r_k,
\end{equation}
with $f_k\ge d^2>0$, then for $|k|< M$
 \begin{equation}\label{p5.2}
 |\tilde x_{k}|\le C\sum_{|j|\le M}e^{-d|k-j|}\tilde r_j+|x_M|e^{-d|M-k|}+
 |x_{-M}|e^{-d|M+k|}.
\end{equation}
\end{proposition}
Notice that Lemma \ref{lem:2} combined with Lemma \ref{lem:4} give us a useful corollary
\begin{corollary}\label{cor:3} For any  function  $\phi(\lambda)$ which has two
bounded derivatives on $[-2+\e,2+\e]$
\begin{equation}\label{phi(J)}
\bigg|\phi(\J^{(n)})_{n+k,n+k}
-\frac{1}{\pi}\int_{-2}^{2}\frac{\phi(\lambda)d\lambda}{\sqrt{4-\lambda^2}}
\bigg|\le C\left(|k|/n\right)^{1/2}+O(n^{-1/3}).
\end{equation}
\end{corollary}
Now let us define a continuous function $q_n(x)$, which for
$x\in\mathbb{Z}/n^{1/3}$ coincides with $x_k$
\[
q_n(\frac{k}{n^{1/3}})=n^{1/3}x_k.
\]
and is a linear function for $x\not\in\mathbb{Z}/n^{1/3}$.  Lemma
\ref{lem:4} allows us to write (\ref{t1.11}) as
\begin{equation}\label{t1.12}
 \frac{
 q_n(\frac{k+1}{n^{1/3}})-2q_n(\frac{k}{n^{1/3}})
 +q_n(\frac{k-1}{n^{1/3}})
 }{n^{-2/3}}=
 2q_n^3\left(\frac{k}{n^{1/3}}\right)+\frac{1}{2P_0(0)}\frac{k}{n^{1/3}}
 q_n\left(\frac{k}{n^{1/3}}\right)
+O(n^{-1/6}),
\end{equation}
where the bound on the remainder is uniform in $|k|/n^{1/3}\le L$
for any $L$.  We are interested in the behaviour of the solution
of this discrete equations which satisfies conditions (cf.
(\ref{l4.1}) and (\ref{as+})):
\begin{equation}\label{t1.12a}
|q_n(x)|\le C|x|^{1/2},\quad |q_n(x)|\le e^{-Cx^{3/2}/2},\quad
as\quad x\to+\infty.
\end{equation}
It follows from Lemma \ref{lem:4} that the functions  $\{q_n(x)\}_{n=1}^\infty$
are uniformly bounded and equicontinuous for any bounded interval. Hence, this family
is weakly compact in any compact set in $\mathbb{R} $ and any convergent
subsequence converges uniformly
to some solution of the Painleve equation (\ref{Pain}), satisfying (\ref{t1.12a}).
Now we need to prove the asymptotic relations (\ref{as-Pain}) for $x\to-\infty$.
 To this aim we use Lemma \ref{lem:5} below, which describes the behavior
of the Stieltjes transform of the following densities (cf (\ref{p_k=})-(\ref{k_n}))
\begin{equation}\label{g}\begin{array}{c}
K_{k,n}(\lambda,\mu)=\sumd_{l=1}^k\psi_l^{(n)}(\lambda)\psi_l^{(n)}(\mu),\\
\rho_{k,n}(\lambda)=n^{-1}K_{k,n}(\lambda,\lambda),\quad

 g_{k,n}(z)=\intd\frac{\rho_{k,n}(\lambda)d\lambda}{z-\lambda}.
\end{array}\end{equation}
\begin{lemma}\label{lem:5} For any $k:|k|\le n^{1/3}\log^2n$
$g_{n+k,n}(z)$ can be represented in the form
\begin{equation}\begin{array}{rcl}
g_{n+k,n}(z)&=&\fracd{1}{2}(V''(0)z+\frac{V^{(4)}(0)}{6}z^3)\\
&&-\fracd{1}{2}X(z)\bigg(P_0^2(0)z^4+\frac{k}{n}P_0(0)z^2+c_k
-\delta_{n+k,n}(z)-\tilde\delta_{n+k,n}(z)\bigg)^{1/2},
\end{array}\label{l5.1}\end{equation}
where $X(z)=\sqrt{z^2-4}$ (here and below we choose the branch which behaves
like $z$ as $z\to+\infty$) and
\begin{equation}
c_k=\pm n^{-5/3}\sum_{j=0}^{|k|} \left(2P_0(0)q_n^2(\frac{\pm
j}{n^{1/3}})\pm\frac{j}{2n^{1/3}}\right)+
n^{-5/6}O\left(\left(|k|/n\right)^{3/2}\right).
\label{c_k}\end{equation}
 ($\pm$ corresponds to the sign of $k$). Moreover, the remainder terms
 $\delta_{n+k,n}(z)$ and $\tilde\delta_{n+k,n}(z)$ in (\ref{l5.1}) for $z:|z|<1$
 admit the bounds
\begin{equation}\begin{array}{rcl}
|\delta_{n+k,n}(z)|&=&\left|n^{-2}\intd
\frac{K_{n+k,n}^2(\lambda_1,\lambda_2)(\lambda_1-\lambda_2)^2}
{(\lambda_1-z)^2(\lambda_2-z)^2}d\lambda_1d\lambda_2\right|\\
&\le&\fracd{C}{n^2|\Im z|^2}\left(1+\frac{|k|}{n|\Im z|^2}+
\frac{1}{n|\Im z|^3}\right)^2,\\
|\tilde\delta_{n+k,n}(z)|&\le& C\bigg[|z|^2(\fracd{k^2}{n^2}+n^{-2/3})+n^{-4/3}
+\frac{\log^{1/2} n}{n^{1/2}}\frac{|z|^4}{|\Im z|}+|z|^5\bigg].
\end{array}\label{l5.2}\end{equation}
\end{lemma}
\noindent The proof of the lemma is given in the next section.

 Let us take $k=-[Ln^{1/3}]$ with $L$ big enough.
Since it is  known (see \cite{HMcL}) that any solution of the Painleve II equations which satisfies
(\ref{t1.12a}) assumes also the bound
\begin{equation}\label{bound}
  q^2(x)\le \frac{-x}{4P_0(0)},\quad x\le -L_0,
\end{equation}
we can conclude that
\begin{equation}\label{as-c_k}
0\le c_k\le\frac{k^2}{4n^2}+O\left(\left(|k|/n\right)^{5/2}\right).
\end{equation}
Now let us choose $\tilde\e=n^{-1/3}P_0^{-1/2}(0)$
and put  in (\ref{l5.1}) $z=\tilde\e\zeta$. Then (\ref{l5.1})
takes the form
\[
g_{n+k,n}(\tilde\e\zeta)=\ti
V(\zeta)-\frac{1}{2}\tilde\e^2P_0(0)X(\tilde\e\zeta)\sqrt{\zeta^4-\zeta^2+
\ti c_k+\tilde \phi(\zeta)},
\]
where $\ti V$ is an analytic function,
\begin{equation}\label{bound1}
0\le\ti c_k=P_0^{-2}(0)\tilde\e^{-4}c_k\le\frac{L^2}{4},
\end{equation}
(see (\ref{bound})), and
\[
\quad
 |\tilde\phi(\zeta)|=P_0^{-2}(0)\tilde\e^{-4}|\delta_{k,n}(\tilde\e\zeta)+\tilde\delta_{k,n}(\tilde\e\zeta)
 +O(kn^{-2})|\le C(1+|\zeta|^2),
\]
 for  $|\Im\zeta|\ge 1$ (see (\ref{l5.2})).  Let
$b$ be the smallest root of the quadratic equation
\[
\zeta^2-L\zeta+\ti c_k=0.
\]
We note, that due to (\ref{bound1}) $b$ is real and positive.
Consider
\begin{multline}
I(b,L)= \fracd{\tilde\e^{-2}}{2\pi i}\oint_{\cal L}
g_{n+k,n}(\tilde\e\zeta)e^{-\zeta^2/2} d\zeta\\ = \fracd{P_0(0)}{2\pi
i}\oint_{\cal L}X(\tilde
\e\zeta)\sqrt{(\zeta^2-b)(\zeta^2-L+b)}e^{-\zeta^2/2} d\zeta +\ti
r_L, \label{I_1}\end{multline} where  ${\cal L}$ consists of two
lines $\Im\zeta=\pm1$ and
\begin{equation}\label{r_L}
|\ti r_L|\le C\oint_{\cal L}\frac{ |\phi(\zeta)|\cdot|X(\tilde
\e\zeta)|
e^{-|\zeta|^2/2}|d\zeta|}{\sqrt{(\zeta^2-b)(\zeta^2-L+b)}}\le
CL^{-1/2}.
\end{equation}
Then,
using the Cauchy theorem, we get
\begin{multline}
I(b,L)= \fracd{P_0(0)}{2\pi }\Im\int \sqrt{((\e x)^2-4)(x^2-
b)(x^2-L+b)} e^{-x^2/2}dx+\ti r_L\\ =\frac{P_0(0)}{2\pi}
\left(-\int_{|x|<b}+\int_{|x|\ge L-b}\right)\sqrt{(x^2-
b)(x^2-L+b)} e^{-x^2/2}  dx +\ti r_L +O(\ti\e)\\=P_0(0)I_1(b,L)
+\ti r_L+O(\ti\e).
\label{p7.6}\end{multline}
One can prove easily
that for large $L$
\[ I_1(b,L)\sim -C_0L^{1/2}b^{3/2},\quad (C_0>0).
\]
On the other hand,
\[
I(b,L)=-\fracd{\tilde\e^{-2}}{2\pi}\int
e^{-x^2/2\sigma}\lim_{\e\to 0}\Im g_{n+k,n}(\tilde\e\zeta)dx>0.
\]
Thus, taking into account (\ref{r_L})
\begin{equation}\label{b-c}
L^{1/2}b^{3/2}\le C'|\ti r_L|\le C''L^{-1/2}\Rightarrow
|c_k|=P_1^2(0)\tilde\e^4(L-b)b/4\le C_1'n^{-4/3}L^{1/3}.
\end{equation}
The last inequality combined with (\ref{c_k}),   and
the bound for the first differences $d^{(1)}_j$ imply for $k=[Ln^{1/3}]$,
$l=[L^{-1/6}n^{1/3}]$
\begin{multline*}
n^{-4/3}O(L^{1/3})=c_{-k}-c_{-k-l}=
\frac{2P_0(0)}{n^{5/3}}\sum_{j=k}^{k+l}
q^2_n(-\frac{j}{n^{1/3}})-\frac{L+L^{-1/6}}{2n^{4/3}}L^{-1/6}\\=
n^{-4/3}\bigg[L^{-1/6}\left((2P_0(0)q^2_n(-\frac{k}{n^{1/3}})
-\frac{L}{2}+O(L^{-1/6})\right)\bigg].
\end{multline*}
Therefore,
\[|q_n(-L)|=(4P_0(0))^{-1/2}L^{1/2}(1+O(L^{-1/2})).
\]
But it is easy to show that any bounded for positive $x$ solution of (\ref{Pain}), which possesses the above
property satisfies also the asymptotic relations
\begin{equation}\label{as-}
q_n(-L)=\tilde s(4P_0(0))^{-1/2}L^{1/2}(1+O(L^{-2})),\quad \tilde s=\hbox{sign}\,q(0).
\end{equation}
Hence, we have proved (\ref{as-Pain}) and now can conclude that
 $q_n(x)$ converge uniformly on any compact in $\mathbb{R}$ to the Hastings-McLeod solution
of (\ref{Pain}), so that
\[
\Delta_n(x)=q_n(x)-q(x)\to 0,\,\, as\,\,\, n\to\infty.
\]
But from (\ref{t1.12}) we derive that for any $x=k/n^{1/3}$ and $h=n^{-1/3}$
we have
\[
h^{-2}(\Delta_n(x+h)+\Delta_n(x-h)-2\Delta_n(x))=
\left[2q_n^2(x)+2q^2(x)+2q_n(x)q(x)+\frac{x}{2P_0(0)}\right]\Delta_n(x)+r(x).
\]
and uniformly in $n$
\[|\Delta_n(x)|\to 0,\,\, as\,\, x\to\pm\infty.
\]
\begin{proposition}\label{p:pain}
For the Hastings - McLeod solution of (\ref{Pain}) there exists
$\delta>0$ such that
\begin{equation}\label{stab}
6q^2(x)+\frac{x}{2P_0(0)}\ge \delta^2.
\end{equation}
\end{proposition}
 This proposition allows us to apply the assertion (\ref{p5.2})
 of Proposition \ref{p:d^2} to $\tilde x_k=\Delta(k/n^{1/3})$ with
 $d=n^{-1/3}\delta$ and $\tilde r_k=r_k-r_k'$ with $r_k$ from (\ref{t1.11})
 and
 \[r'_k=q(\frac{k+1}{n^{1/3}})-2q(\frac{k}{n^{1/3}})
 +q(\frac{k-1}{n^{1/3}}) -
 2n^{-2/3}q^3(\frac{k}{n^{1/3}})+\frac{4}{2nP_0(0)}
 q(\frac{k}{n^{1/3}})=O(n^{-1})\]
 uniformly in $k$.  The assertion of  Theorem \ref{thm:1} follows.

\medskip

{\it Proof of Theorem \ref{thm:2}.} Take some fixed
$\zeta_1$, $\zeta_2$ with $\Im\zeta_{1,2}\not=0$, denote
$z_{1,2}=\zeta_{1,2}n^{-1/3}$ and consider the functions:
\begin{equation}\begin{array}{rcl}
F_{n}(\zeta _{1},\zeta _{2})&=&\intd (\lambda _{1}-z_{1})^{-1}
(\lambda _{2}-z_{2})^{-1}(\lambda_1-\lambda_2)^2K_{n}^{2}(\lambda
_{1},\lambda _{2})d\lambda _{1}d\lambda _{2},\\ F_{n}^{(1)}(\zeta
_{1})&=&n^{-2/3}\intd (\lambda _{1}-z_{1})^{-2}(\lambda
_{2}-z_{1})^{-2} (\lambda_1-\lambda_2)^2K_{n}^{2}(\lambda
_{1},\lambda _{2})d\lambda _{1}d\lambda _{2}.
\end{array}\label{F(z_1,z_2)}\end{equation}
 Changing variables
$\lambda_{1,2}=t_{1,2}n^{-1/3}$, and using
(\ref{def-K}), we get
\begin{equation}\label{F-1}\begin{array}{rcl}
F_{n}(\zeta _{1},\zeta _{2})&=&\intd  (t
_{1}-\zeta_{1})^{-1}(t_{2}-\zeta_{2})^{-1}(t_1-t_2)^2{\cal
K}_{n}^{2}(t_{1},t_{2})dt_{1}dt_{2}.
\end{array}\end{equation}
The proof of Theorem \ref{thm:2} is based on the following proposition:
\begin{proposition}\label{pro:w_conv}
Let the functions $F_n$ and $F_n^{(1)}$ be defined by (\ref{F(z_1,z_2)})
and there exists $F(\zeta _{1},\zeta _{2})$ of the form
\[
F(\zeta _{1},\zeta _{2})=\int \int  (t
_{1}-\zeta_{1})^{-1}(t_{2}-\zeta_{2})^{-1}
(t_1-t_2)^2\Phi(t_1,t_2)dt_{1}dt_{2}
\]
with $\Phi(t_1,t_2)$ bounded uniformly in each compact
in $\mathbb{R}^2$ and   such that  uniformly in $
\Im\zeta_{1,2}\ge 1$
\begin{equation}\label{t2.0}
  |F_{n}(\zeta _{1},\zeta _{2})-F(\zeta _{1},\zeta _{2})|\le C (1+|\zeta|^2)n^{-1/4}.
\end{equation}
Let  also uniformly in $Im \zeta\ge\varepsilon_n=(\log n)^{-1/2}$  and
 $\Re\zeta$ varying in any compact  $L\subset\mathbb{R}$
\begin{equation}\label{t2.0a}
 |F_{n}^{(1)}(\zeta) |\le C(L).
\end{equation}
Then for any intervals $I_1,I_2\subset\mathbb{R}$
\[
\lim_{n\to\infty}\int_{I_1}dt_1\int_{I_2}dt_2 {\cal K}_{n}^{2}(t_1,t_2)=
\int_{I_1}dt_1\int_{I_2}dt_2 \Phi(t_1,t_2).
\]
\end{proposition}
\noindent{\it Proof of Proposition \ref{pro:w_conv}.}
Notice that
\[F_{n}^{(1)}(\zeta)=n^{4/3}\delta_{n,n}(z),\]
where $\delta_{k,n}(z)$ is defined in (\ref{l5.2}). Therefore, using the bound
(\ref{t2.0a}) in (\ref{l5.1}), we get for any $\Im\zeta\ge\varepsilon_n$
\begin{equation}\label{p6.1}
 | g_{n,n}(n^{-1/3}\zeta)+g_{n,n}(-n^{-1/3}\zeta)|\le Cn^{-2/3}(|\zeta|^2+1),
\end{equation}
where $C$ does not depend on $n$ and $\zeta$.
On the other hand, taking $z=n^{-1/3}(a+i\varepsilon)$, we have for any $\epsilon>\epsilon_n$
\begin{multline}\label{p6.2}
\int_{|t-a|\le\varepsilon}{\cal K}_n(t,t)dt\le
2\varepsilon^2 \int \frac{{\cal K}_n(t,t)dt}{(t-a)^2+\varepsilon^2}
=2\varepsilon n^{2/3}\Im g_{n,n}(z)\\ \le\varepsilon_n n^{2/3}(\Im g_{n,n}(z)
+\Im g_{n,n}(-z))\le\varepsilon C(a^2+1).
\end{multline}
Take the integral
\[\int_{\Im\zeta_1=\pm 1}d\zeta_1\int_{\Im\zeta_2=\pm 1}d\zeta_2
(F_{n}(\zeta _{1},\zeta _{2})-F(\zeta _{1},\zeta _{2}))e^{-(\zeta_1-a_1)^2/2\sigma_1}
e^{-(\zeta_2-a_2)^2/2\sigma_2}.
\]
Using the Cauchy theorem, we get that
 for any $\sigma_{1,2}>0$, $a_{1,2}\in\mathbb{R}$
 \[
\bigg|\int \int (t_1-t_2)^2\left({\cal K}_{n}^{2}(t
_{1},t_{2})-\Phi(t_1,t_2)\right)e^{-(t_1-a_1)^2/2\sigma_1}
e^{-(t_2-a_2)^2/2\sigma_2} dt_{1}dt_{2}\bigg|\le C n^{-1/4}
 \]
with $C$, depending on $a_1,a_2,\sigma_1,\sigma_2$, but independent of $n$. This
implies that for any Lipshitz $f_1$ and $f_2$ with a compact support
 \begin{equation}\label{}
 \lim_{n\to\infty}\int(t_1-t_2)^2\left({\cal K}_{n}^{2}(t_{1},t_{2})-
 \Phi(t_1,t_2)\right)f_1(t_1)f_2(t_2)dt_1dt_2=0.
 \end{equation}
 For any  small enough $\epsilon$ denote by $f_{1}^{(+\epsilon)}$ a
 Lipshitz function which  coincides with the indicator
 $\chi_{I_1}$  of $I_1=(a_1,b_1)$ inside  this interval,
 equals to zero  outside of $(a_1-\epsilon,b_1+\epsilon)$ and is
 linear in $(a_1-\epsilon,a_1),(b_1,b_1+\epsilon)$. Let $f_{1}^{(-\epsilon)}$
 be a similar function   for the interval $(a_1+\epsilon,b_1-\epsilon)$ and
 $f_{2}^{(\pm\epsilon)}$ be  similar functions for $I_2$. Denote also
 \[\phi_{\varepsilon_1}(t_1,t_2)=(t_1-t_2)^{-2}\mathbf{1}_{|t_1-t_2|>\varepsilon_1}+
\varepsilon_1^{-2}\mathbf{1}_{|t_1-t_2|\le\varepsilon_1}. \]
 Then, evidently
 \[ \phi_{\varepsilon_1}(t_1,t_2) f_{1}^{(-\epsilon)}(t_1)f_{2}^{(-\epsilon)}(t_2)\le
 \phi_{\varepsilon}(t_1,t_2)\chi_{I_1}(t_1)\chi_{I_1}(t_2)
 \le\phi_{\varepsilon_1}(t_1,t_2) f_{1}^{(+\epsilon)}(t_1)f_{2}^{(+\epsilon)}(t_2).
 \]
Integrate this inequality with $(t_1-t_2)^2{\cal K}_{n}^{2}(t_{1},t_{2})$,
 and take
the limits $n\to\infty$ and then $\epsilon\to 0$. We obtain
\begin{multline*}\int_{I_1\times I_2}dt_1dt_2\Phi(t_1,t_2)-O(\varepsilon_1)\le
\lim_{n\to\infty}\int_{I_1\times I_2}dt_1dt_2{\cal K}_n^2(t_1,t_2)\\+
\lim_{n\to\infty}\int_{I_1\times I_2}dt_1dt_2{\cal K}_n^2(t_1,t_2)\left(
\varepsilon_1^{-2}(t_1-t_2)^2-1\right)
\mathbf{1}_{|t_1-t_2|\le\varepsilon_1}\le
\int_{I_1\times I_2}dt_1dt_2\Phi(t_1,t_2)+O(\varepsilon_1).
\end{multline*}
But using the inequality  ${\cal K}_n^2(t_1,t_2)\le
{\cal K}_n(t_1,t_1){\cal K}_n(t_2,t_2)$ and then (\ref{p6.2}), we obtain that
the second limit is $O(\varepsilon_1)$. Then, taking
the limit $\varepsilon_1\to 0$ we get the assertion of Proposition \ref{pro:w_conv}.

\medskip

Let us check that in our case conditions (\ref{t2.0})
and (\ref{t2.0a}) are satisfied. Using the Christoffel-Darboux formula, it is easy to derive from
(\ref{F(z_1,z_2)}) that
\begin{equation}\begin{array}{rcl}
F_{n}(\zeta _{1},\zeta _{2})&=& (J_{n}^{(n)})^2\intd
\frac{(\psi^{(n)}_n(\lambda_1)\psi^{(n)}_{n+1}(\lambda_2)-
\psi^{(n)}_n(\lambda_2)\psi^{(n)}_{n+1}(\lambda_1))^2}{(\lambda
_{1}-z_{1}) (\lambda _{2}-z_{2})}
\lambda _{1}d\lambda_{2}\\
&=&(J_{n}^{(n)})^2[R_{n,n}(z_1)R_{n+1,n+1}(z_2)+ R_{n,n}(z_2)
R_{n+1,n+1}(z_1)\\
&&{\hskip 4cm}-2 R_{n,n+1}(z_1)R_{n-1,n}(z_2)],\\
F_{n}^{(1)}(\zeta _{1})&=&2(J_{n}^{(n)})^2[R'_{n,n}(z_1)R'_{n+1,n+1}(z_1)
-R'_{n,n+1}(z_1)R'_{n,n+1}(z_1)],
\end{array}\label{F-2}\end{equation}
where
\begin{equation}
R_{k,m}(z)=\int\frac{\psi^{(n)}_{k}(\lambda)\psi^{(n)}_{m}(\lambda)}
{z-\lambda}d\lambda
 \label{R_kk}
\end{equation}
is the resolvent of $J^{(n)}$ ($R=(z-J^{(n)})^{-1}$).
\begin{proposition}\label{pro:res}
Let $\mathcal{J}$ be an arbitrary Jacobi matrix , with
$|\mathcal{J}_{j,j+1}|\le A$, for all $j$ such that $|j-k|\le M$.
Consider  $\mathcal{R}(z)=(z-\mathcal{J})^{-1}$ with $|\Im z|\le
A_1$. Then
\begin{equation}
|\mathcal{ R}_{k,j}(z)|\leq \frac{C_{1}'}{\Im z}e^{-C_{2}'|\Im z|
|k-j|}+\frac{C_{1}'}{|\Im z|^2}e^{-C_{2}'|\Im z|M},
\label{exp_b_R}
\end{equation}
where $C_{1}',C_{2}'>0$ depend only on $A$ and $A_1$.
\end{proposition}
Let us choose $N=[n^{1/3}\log ^2n ]$. By (\ref{exp_b_R}), for
$\alpha=0,1$ and $\Im z>n^{-1/3}\varepsilon_n$
\[\sum_{k:|k-n|>N}|R_{n-\alpha,k}(z)|^2\le Cne^{-CN\Im z}/\Im z
\le e^{-C_1\log^{3/2}n}.\]
Therefore for $\alpha,\beta=0,1$
\begin{equation}\label{t2.1a}
R'_{n+\alpha,n+\beta}(z)=\sum_{|k|<N}
R_{n+\alpha,k}(z)R_{n+\beta,k}(z) +O(e^{-C\log^{3/2}n}).
\end{equation}
Consider the matrix
 $\J^{(n,2N)}$ whose entries coincide with
that of $\J^{(n)}$ with the only exception $J^{(n,2N)}_{n\pm
2N,n\pm 2N+1}=0$. Consider
\[
R^{(2N)}(z)=(z-\J^{(n,2N)})^{-1},
\]
We need also a simple observation, following from
 (\ref{exp_b_R}) and the resolvent identity
\begin{equation}\label{res-id}
R^{(1)}-R^{(0)}=R^{(0)}(M^{(1)}-M^{(0)})R^{(1)},\quad
R^{(1,0)}=(z-M^{(1,0)})^{-1}.
\end{equation}
\begin{remark}\label{rem:8}
If in the resolvent identity we take
 $M^{(0)}=\J^{(n)}$ and $M^{(1)}=\J^{(n,2N)}$ with $N_k=|k|+n^{1/3}\log^2n$, then
 (\ref{exp_b_R}) gives
us for any $z:|\Im z|>n^{-1/3}\log^{-1/2}n$ for any fixed $\delta$
\begin{equation}\label{t2.1}
|R_{n+k,n+k+\delta}(z) -R^{(2N)}_{n+k,n+k+\delta}|\le C_1'
e^{-C_2'\log^{3/2}n}.
\end{equation}
\end{remark}
Let us study first the case when in (\ref{t1.1}) $\tilde s=1$.
Consider the Dirac operator $\mathcal{A}$ defined in
$L_2(\mathbb{R})\times L_2(\mathbb{R})$ by the differential
expression (\ref{A})-(\ref{as-Pain}). Let ${\cal
R}_{\alpha,\beta}(x,y;\zeta)$  $(\alpha,\beta =0,1)$ be the kernel
of the operator ${\cal R}(\zeta)=(\zeta-2{\cal A})^{-1}$. It means
that the coefficients ${\cal R}_{\alpha,\beta}(x,y;\zeta)$ satisfy
the equations
\begin{equation}\label{res}\begin{array}{rcl}
  2\fracd{d}{dx}{\cal R}_{1,0}(x,y;\zeta)+2q(x){\cal R}_{1,0}(x,y;\zeta)-
  \zeta {\cal R}_{0,0}(x,y;\zeta)  & = &   \delta(x-y) \\
  - 2\fracd{d}{dx}{\cal R}_{0,1}(x,y;\zeta)+2q(x){\cal R}_{0,1}(x,y;\zeta)
  -\zeta {\cal R}_{1,1}(x,y;\zeta)  & = &  \delta(x-y)\\
  - 2\fracd{d}{dx}{\cal R}_{0,0}(x,y;\zeta)+2q(x){\cal R}_{0,0}(x,y;\zeta)
  -\zeta {\cal R}_{1,0}(x,y;\zeta)  & = & 0 \\
2\fracd{d}{dx}{\cal R}_{1,1}(x,y;\zeta)+2q(x){\cal
R}_{1,1}(x,y;\zeta) -\zeta {\cal R}_{0,1}(x,y;\zeta)  & =
&0
\end{array}\end{equation}
Here  $\delta(x)$ is the Dirac $\delta$-function and, e.g., the first equation means that
the l.h.s. is equal to zero, as $x\not=y$ and
$2{\cal R}_{1,0}(x+0,x)-2{\cal R}_{1,0}(x-0,x)=1$.

Consider the $(4N+1)\times (4N+1)$ matrix with entries
\begin{equation}\label{R^*}
R^*_{n+2k+\alpha,n+2m+\beta}=(-1)^{(k+m)}{\cal R}_{\alpha,\beta}
\left(\frac{2k+\alpha}{n^{1/3}},\frac{2m+\beta}{n^{1/3}};\zeta\right),
\end{equation}
where $ -N\le k,m\le N-1,\quad \alpha,\beta=0,1 $. Let us prove,
that for $|k|,|m|\le N$
\begin{equation}\label{t2.2}
\left|R^{(2N)}_{n+k,n+m}(\zeta n^{-1/3})-R^*_{n+k,n+m}\right|\le
Cn^{-1/4}(1+|\zeta|^2).
\end{equation}
Using  equations (\ref{res}), one can check  that for all
$ -2N\le k,m\le 2N,
\quad \alpha,\beta=0,1$
\begin{equation}\label{t2.3}
  ((z-J^{(n,2N)})R^*)_{n+k,n+m}=\delta_{k,m}+d_{k,m}+a^{(1)}_m\delta_{k,2N}+
  a^{(2)}_m\delta_{k,-2N}
\end{equation}
where the remainder terms could be estimated as follows:
\begin{equation}\label{t2.4}\begin{array}{rcl}
|d_{k,m}|&\le &Cn^{-2/3}\sumd_{\alpha,\beta=0,1}
\left(1+|\zeta|^2+q^2(x)\right)
|{\cal R}_{\alpha,\beta}
(x,y;\zeta)|\bigg|_{x=\frac{k+\alpha}{n^{1/3}},
y=\frac{m+\theta(\alpha,\beta,k,m)}{n^{1/3}}}\\
|d_{k,k}|&\le &Cn^{-1/3}\sumd_{\alpha,\beta=0,1}(1+|\zeta|+q(x))|{\cal R}_{\alpha,\beta}
(x,y;\zeta)|\bigg|_{x=\frac{k+\alpha}{n^{1/3}},
y=\frac{k+\theta(\alpha,\beta,k,k)}{n^{1/3}}}\end{array}\end{equation}
with $|\theta_{\alpha,\beta,k,m}|\le 2$.
The components of the vectors ${\bf a}^{(1)}$ and ${\bf a}^{(2)}$ satisfy the bounds
\begin{equation}\label{t2.4a}\begin{array}{rcl}
|a^{(1)}_{2m+\alpha}|&\le&
|{\cal R}_{0,\alpha}((2N+1)n^{-1/3},(2m+\alpha)n^{-1/3};\zeta)|+
Cn^{-1/3}(|\zeta|+|\log^2n|)
\\
|a^{(2)}_{2m+\alpha}|&\le &
|{\cal R}_{0,\alpha}((-2N-1)n^{-1/3},(2m+\alpha)n^{-1/3};\zeta)|+
Cn^{-1/3}(|\zeta|+|\log^2n|).
\end{array}\end{equation}
Let us define the $(4N+1)\times(4N+1)$ matrices
\[
D_{n+i,n+j}=d_{n+i,n+j},\quad \tilde D_{n+i,n+j}=a^{(1)}_j\delta_{i,2N}+
 a^{(2)}_j \delta_{i,-2N}.
\]
Then (\ref{t2.3}) can be rewritten as
\begin{equation}\label{t2.5}
(z-J^{(n,2N)})R^*=I+D+\tilde D.
\end{equation}
Using the trivial bound for the norm of any matrix $||A||^2\le \hbox{Tr}AA^*$
 and the bound  for the resolvent of the Dirac operator (see \cite{LS})
 \[\int|\mathcal{R}_{\alpha,\beta}(x,y;\zeta)|^2dy\le
 C|\Im\zeta|^{-2}(|q(x)|+|\zeta|),\]
we obtain from the first and the second line of (\ref{t2.4}) that for
$|\Im\zeta|>\varepsilon_n$
\begin{equation}\label{||D||}
  ||D||\le C (|\zeta|+1)n^{-1/3}\log^3n\le C (|\zeta|+1)n^{-1/4}.
\end{equation}
On the other hand, it  follows from (\ref{t2.5}) that for any $-2N\le k,m\le 2N$
\begin{equation}\label{t2.6}
R^{(2N)}_{n+k,n+m}=(z-J^{(n,2N)})^{-1}_{n+k,n+m}=(R^*(I+D+\tilde D)^{-1})_{n+k,n+m}
\end{equation}
and so we can write
\begin{equation}\label{t2.7}
|R^{(2N)}_{n,n}-(R^*(I+\tilde D)^{-1})_{n,n}|\le ||D||\cdot||R^*||
\frac{||(I+\tilde D)^{-2}||}{1-||D||\,||(I+\tilde D)^{-2}||}.
\end{equation}
But  for any ${\bf x}\in\mathbb{R}^{4N+1}$,
\begin{equation}\label{I+D}
(I+\tilde D){\bf x}={\bf x}+{\bf e}_{n+2N}({\bf x},{\bf a}^{(1)})
+{\bf e}_{n-2N}({\bf x},{\bf a}^{(2)}),
\end{equation}
where $\{{\bf e}_k\}_{k=n-2N}^{n+2N}$ is a standard basis in $\mathbb{R}^{4N+1}$.
So we get
\begin{equation}\label{t2.7a}
||(I+\tilde D)^{-1}||\le ||A^{-1}||+1,\quad A=\left(
\begin{array}{cc}1+a^{(1)}_{n+2N-1}&a^{(1)}_{n-2N}\\
a^{(2)}_{n+2N-1}&1+a^{(2)}_{n-2N}
\end{array}\right).
\end{equation}
The off diagonal entries of $A$ tend to zero, because they are
expressed in terms of the resolvent $\mathcal{
R}_{\alpha,\beta}(x,y;\zeta)$ with $|x-y|\sim\log^2 n$ (see
(\ref{t2.4a})). $a^{(1)}_{n-2N-1}\to 0$, because $q(x)\sim
C\sqrt{-x}$, as $x\to-\infty$. And $|a^{(2)}_{n+2N}|\to 1/4$,
because if the potential tends to zero fast enough, then the
resolvent near the diagonal coincides asymptotically  with that
for $q=0$ (see \cite{LS}). Hence, using (\ref{t2.7}) and
(\ref{t2.7a}), we get
\begin{equation}\label{t2.8}
\left|R^{(2N)}_{k,m}-(R^*(I+\tilde D)^{-1})_{k,m}\right|\le Cn^{-1/4}(1+|\zeta|^2).
\end{equation}
But from (\ref{I+D}) we derive
 \[
(I+\tilde D)^{-1}_{k,m}=\delta_{k,m},\quad k\not=\pm 2N.
\]
Hence, we obtain from (\ref{t2.8}) and (\ref{exp_b_R})
\begin{equation}\label{t2.9}
|R^{(2N)}_{n,n}-R^*_{n,n}|\le Cn^{-1/4}+||(I+\tilde D)^{-1}||(|R^*_{n,2N}|+
|R^*_{n,-2N}|)\le Cn^{-1/4}
\end{equation}
and we have proved (\ref{t2.0}) with
\begin{multline}
F(\zeta _{1},\zeta _{2})=
( {\cal R}_{0,0}(0,0;\zeta_1){\cal R}_{1,1}(0,0;\zeta_2)
+{\cal R}_{0,0}(0,0;\zeta_2)
{\cal R}_{1,1}(0,0,\zeta_1)
\\-2 {\cal R}_{0,1}(0,0+0;\zeta_1){\cal R}_{1,0}(0+0,0;\zeta_2)),
\label{F-3}
\end{multline}
where we denote ${\cal R}_{0,1}(0,0+0,\zeta_1)=\lim_{x\to +0}{\cal R}_{0,1}(0,x,\zeta_1)$.
But according to the spectral theorem (see \cite{LS}),
\begin{equation}\label{res_D}
\mathcal{R}_{\alpha,\beta}(x,y;\zeta)=\int\frac{\Psi_\alpha(x; t)
\Psi_\beta(y; t)}{\zeta-2t}dt,
\end{equation}
where $\mathbf{\Psi}(x;t)=(\Psi_0(0;t),\Psi_1(0;t))$ is the
solution of the Dirac system (\ref{A}), satisfying asymptotic
conditions (\ref{as-Psi}). The last two relations and the formula
of the inverse Stieltjes transform yield
\begin{equation}\label{t2.10}
\Phi(t_1,t_2)=(2\pi)^{-2}\left(
\Psi_1(0; t_1/2)\Psi_0(0; t_2/2)-
\Psi_0(0; t_2/2)\Psi_1(0; t/2)\right)^2.
\end{equation}
Moreover, similarly to (\ref{t2.9}) we obtain
\begin{multline}\label{t2.9a}
n^{-1/3}\sum_{|k-n|\le N} (R^{(2N)}_{n,k})^2\le
\int \mathcal{R}^{2}_{0,0}(0,y;\zeta)dy+
\int \mathcal{R}^{2}_{0,1}(0,y;\zeta)dy+
O(||D||\cdot|\Im\zeta|^{-2})\\
=-\frac{\partial}{\partial\zeta}\mathcal{R}_{0,0}(0,0;\zeta)+O(||D||\cdot|\Im\zeta|^{-2}).
\end{multline}
Using the representation (\ref{res_D}) and taking into account that
$\Psi_\alpha(x; t)$  are smooth function with respect to $t$, according
to the standard
theory of the Cauchy type integrals (see \cite{Mus}) we get that the
derivative in the r.h.s. of
(\ref{t2.9a}) is uniformly bounded up to the real line. Therefore we obtain
(\ref{t2.0a}) and then, on the basis of Proposition \ref{pro:w_conv}), obtain
the assertion of Theorem \ref{thm:2}
for $l=2$. For the other $l$ we  study by the same way
\begin{multline*}
F_{n}(\zeta _{1},\dots,\zeta_l)=\int\prod_{i=1}^{l} (t_{i}-\zeta_{i})^{-1}
(t_1-t_2)\dots(t_l-t_1){\cal K}_{n}(t_{1},t_{2})\dots
{\cal K}_{n}(t_{l},t_{1}) dt_{1}\dots dt_{l}
\end{multline*}
 Now, notice that the $(\Psi_0(x,t),\Psi_1(x,t))\to(-\Psi_1(x,t),\Psi_0(x,t))$
gives us the solution of (\ref{A}) with potential $q_1(x)=-q(x)$ but does not change
the expression (\ref{exp_K}). This completes
 the proof of Theorem \ref{thm:2}

 To prove Corollary 1 we split the expansion for the Fredholm determinant in two parts:
 with $m<N$ and $m\ge N$ ($m$ is the number of variables in the correspondent
 determinant). Using the Hadamard bound for determinants with $m>N$ and then (\ref{p6.2})
it is easy to see that the second sum possesses the bound $C^N/N!$. Hence using Theorem
\ref{thm:2} we can take the limit $n\to\infty$ in the first sum and then take the limit $N\to
\infty$.   Relation (\ref{hole}) follows.

\section{Auxiliary results}\label{sec:3}

{\it Proof of Lemma \ref{lem:1}.}
 We introduce an eigenvalue
distribution which is more general than (\ref{p(la)}),
making different the number of variable and the large parameter in front of $%
V$ in the exponent of the r.h.s of (\ref{p(la)}):
\begin{equation}
p_{k,n}(\lambda _{1},...\lambda _{k})=Z_{k,n}^{-1}\prod_{1\leq
j<m\leq k}(\lambda _{j}-\lambda _{m})^{2}\exp
\prod_{j=1}^{k}e^{-nV(\lambda _{j})}, \label{p^k(la)}
\end{equation}
where $Z_{k,n}$ is the normalizing factor. For $k=n$ this
probability distribution density coincides with (\ref{p(la)}). Let
\begin{equation}\label{2.2'}
\displaystyle{\ \tilde{\rho}_{k,n}(\lambda _{1})=\int d\lambda
_{2}...d\lambda _{k}p_{k,n}(\lambda _{1},...\lambda _{k})}, \quad
\displaystyle{\ \tilde{\rho}_{k,n}(\lambda _{1},\lambda _{2})=\int
d\lambda _{3}...d\lambda _{k}p_{k,n}(\lambda _{1},...\lambda
_{k})}
\end{equation}
be the first and the second marginal densities of (\ref{p^k(la)}).
By the standard argument \cite{Me:91} we obtain
\begin{equation}\begin{array}{l}
\tilde{\rho}_{k,n}(\lambda )=k^{-1}{K}_{k,n}(\lambda ,\lambda ),\\
\displaystyle{
 \tilde{\rho}_{k,n}(\lambda ,\mu
)={\frac{1}{k(k-1)}}[{K}_{k,n}(\lambda,\lambda){K}_{k,n}(\mu,\mu)
-{K}_{k,n}^2(\lambda,\mu)],}
\end{array}\label{rho-K}
\end{equation}
where ${K}_{k,n}(\lambda ,\mu )$ is defined in (\ref{g}).
Remark also  that
\[
\tilde{\rho}_{k,n}(\lambda )=\frac{n}{k}\rho _{k,n}(\lambda ),
\]
where $\rho _{k,n}$ is defined in (\ref{g}). Taking any  twice
differentiable  and vanishing outside $\sigma_{2\e}$  function
$\phi(\lambda)$ and integrating by parts with respect to $V$, we
come to the identity
\begin{equation}
\int V^{\prime }(\lambda )\tilde{\rho}_{k,n}(\lambda)
\phi(\lambda) d\lambda ={\frac{1}{n}}\int
\tilde{\rho}_{k,n}(\lambda )\phi'(\lambda)d\lambda
+2{\frac{k-1}{n}}\int \tilde{\rho}_{k,n}(\lambda ,\mu)
 \frac{\phi(\lambda)}{\lambda-\mu}d\lambda
d\mu .
\label{equrho1}\end{equation}
The symmetry property $\tilde{\rho}_{k,n}(\lambda ,\mu )=\tilde{\rho%
}_{k,n}(\mu ,\lambda )$ of (\ref{2.2'}) implies
\[
\int \tilde{\rho}_{k,n}(\lambda ,\mu )\frac{\phi(\lambda)}{\lambda
-\mu } d\lambda d\mu =-\int\tilde{\rho}_{k,n}(\lambda ,\mu)
\frac{\phi(\mu)}{\lambda -\mu }d\lambda d\mu.
\]
This allows us to rewrite (\ref{equrho1}) in the form
\[
\int V^{\prime }(\lambda )\tilde{\rho}_{k,n}(\lambda
)\phi(\lambda) d\lambda =\frac{1}{n}\int
\tilde{\rho}_{k,n}(\lambda )\phi'(\lambda) d\lambda
+\frac{k-1}{n}\int \tilde{\rho}_{k,n}(\lambda ,\mu )
\frac{\phi(\lambda )-\phi(\mu )}{\lambda-\mu}d\lambda d\mu .
\]
Now, using (\ref{rho-K}) and the fact that
\[
\int K_{k,n}^2(\lambda,\mu)d\mu=K_{k,n}(\lambda,\lambda),
\]
we can rewrite the last equation as
\begin{equation}
\int\frac{\phi(\lambda)-\phi(\mu)}{\lambda-\mu}\rho_{k,n}(\lambda)
\rho_{k,n}(\mu)d\lambda d\mu-\int
V'(\lambda)\rho_{k,n}(\lambda)\phi(\lambda)d\lambda
+\delta_{k,n}(\phi)=0,
 \label{equg}
\end{equation}
where we denote
\begin{equation}\begin{array}{l}
\delta_{k,n}(\phi)=\fracd{1}{2n^{2}}\int\bigg(\phi'(\lambda)+
\phi'(\mu)-2\frac{\phi(\lambda)-\phi(\mu)}{\lambda-\mu}\bigg)
K_{k,n}^2(\lambda,\mu)d\lambda d\mu.
\end{array}\label{delta}\end{equation}
 Subtracting from (\ref{equg}) the
relation obtained from (\ref{equg}) by the replacement $k\to(k-1)$
and multiplying the difference by $n$, we obtain:
\begin{multline}
2\int\frac{\phi(\lambda)-\phi(\mu)}{\lambda-\mu}
\rho(\mu)[\psi^{(n)}_{k}(\lambda)]^2d\lambda d\mu -\int
V'(\lambda)\phi(\lambda)[\psi^{(n)}_{k}(\lambda)]^2d\lambda\\
+\delta_{k,n}^{(R)}(\phi)+\tilde\delta_{k,n}^{(R)}(\phi)
=0,
 \label{equR}
\end{multline}
where
\begin{equation}\begin{array}{rcl}
\delta_{k,n}^{(R)}(\phi)&=& \fracd{1}{n}\int
K_{k,n}(\lambda,\mu)\psi^{(n)}_{k}(\lambda)\psi^{(n)}_{k}(\mu)
\bigg(\phi'(\lambda)+
\phi'(\mu)-2\frac{\phi(\lambda)-\phi(\mu)}{\lambda-\mu}\bigg)d\lambda
 d\mu
 \\
\tilde\delta_{k,n}^{(R)}(\phi)&=&\intd\frac{\phi(\lambda)-\phi(\mu)}{\lambda-\mu}
(\rho_{k,n}(\mu)-\rho(\mu))[\psi^{(n)}_{k}(\lambda)]^2d\lambda
d\mu-\frac{1}{n}\int\phi'(\lambda)[\psi^{(n)}_{k}(\lambda)]^2d\lambda.
 \end{array} \label{de^R}
\end{equation}
By  Schwartz inequality
\begin{equation}\begin{array}{l}
|\delta_{k,n}^{(R)}(\phi)|\le \fracd{2}{n}||\phi''||_0\bigg(\int
K_{k,n}^2(\lambda,\mu)(\lambda-\mu)^2d\lambda
 d\mu\bigg)^{1/2}\\{\hskip 3cm}\cdot\bigg(\intd(\psi^{(n)}_{k}(\lambda))^2(\psi^{(n)}_{k}(\mu))^2d\lambda
d\mu\bigg)^{1/2}
\le\frac{C}{n}||\phi''||_0\\
|\tilde\delta_{k,n}^{(R)}(\phi)|\le|\tilde\delta_{0,n}^{(R)}(\phi)|+
\fracd{|k-n|}{n}||\phi'||_0\le C\bigg(||\phi'||_2^{1/2}||\phi''||_2^{1/2}
\frac{\log^{1/2}n}{n^{1/2}}+
||\phi'||_0\fracd{|k-n|}{n}\bigg),
 \end{array} \label{b_de^R}\end{equation}
 where the symbols $||...||_0$ and $||...||_2$ denotes the supremum and the $L_2$-norm on
$\sigma_\varepsilon$.
 Here we have used
 the result of \cite{BPS}, valid for any smooth
  function $\phi(\mu)$ defined on $\sigma_\varepsilon$
\begin{equation}
\left|\int\phi(\mu)\rho_{n,n}(\mu)d\mu-\int\phi(\mu)\rho(\mu)d\mu\right|\le
C||\phi'||_2^{1/2}||\phi||_2^{1/2}n^{-1/2}\log^{1/2}n,
\label{univ.1}
\end{equation}
where the symbol $||...||_2$ denotes the  $L_2$-norm on
$\sigma_\varepsilon$.

Now we are going to use (\ref{rep-V}) in the second integral in the r.h.s. of
(\ref{equR}). But since this representation is valid only for $\lambda\in [-2,2]$
we need to restrict the integrals in  (\ref{equR}) by some
$\sigma_{\tilde\e}=[-2-\tilde\e,2+\tilde\e]$ with  some small $\tilde\e>0$.
To  this aim we use
\begin{proposition}
\label{pro:1} Consider any unitary invariant ensemble of the form
(\ref{p(M)}) and assume that  $V(\lambda )$ possess
two bounded derivatives in some neighborhood of the support
$\sigma $ of the density of states $\rho $. Let also $\sigma $ consist of
a finite number of  intervals,   $\rho(\lambda)$
satisfy condition C4 and $\rho(\lambda)\sim C(a^*)|\lambda-a^*|^{1/2}$
near any edge point $a^*$ of $\sigma $.

 Then there exist absolute constants
 $C,C_{0},\varepsilon _{0}>0$ such that for any positive
 $C_0n^{-1/2}\log n\le\varepsilon \le\varepsilon _{0}$ and for any integer
 $k:|k|\le n+n^{1/2}$  the bounds hold:
\begin{equation}
\int_{\mathbf{R}\setminus \sigma _{\varepsilon }}\rho
_{k,n}(\lambda )d\lambda \leq e^{-nC\varepsilon },\quad
\int_{\mathbf{R}\setminus \sigma _{\varepsilon }}(\psi
_{k}^{(n)}(\lambda ))^{2}d\lambda \leq e^{-nC\varepsilon }.
\label{tails}
\end{equation}
\end{proposition}

This proposition was proved in \cite{APS2}. It allows us to restrict the integration
in the first three integrals of (\ref{equR}) by $\sigma_{\tilde\e}$ with
$\tilde\e=C_0n^{-1/2}\log n$. Now we can use (\ref{rep-V}).
The error, which appear because of this replacement is
of the order $O(\tilde\e)$, because $V'(\lambda)$ is a smooth
function in $\sigma_{\tilde\e}$.
Hence, (\ref{equR}) can be rewritten in the form
\begin{equation}
2\int_{\sigma_{\tilde\e}}(\psi^{(n)}_{k}(\lambda))^2d\lambda
\int_{-2}^{2}\frac{\phi(\mu)}{\lambda-\mu}
\rho(\mu) d\mu
=\delta_{k,n}^{(R)}(\phi)+\tilde\delta_{k,n}^{(R)}(\phi)+O(||\phi||_0n^{-1/2}\log n).
 \label{equR1}
\end{equation}
Take $\phi(\lambda)=P_0^{-1}(\lambda)(z-\lambda)^{-1}$ and
substitute in (\ref{equR1}). Then, according to (\ref{rho}), we
get
\begin{equation}
2\int_{\sigma_{\tilde\e}}(\psi^{(n)}_{k}(\lambda))^2d\lambda\int_{-2}^2
\frac{\mu^2\sqrt{4-\mu^2}}{(z-\mu)(\lambda-\mu)}
 d\mu =\delta_{k,n}^{(R)}(z)+\tilde\delta_{k,n}^{(R)}(z)
 +O(|\Im z|^{-1}n^{-1/2}\log n),
 \label{equR2}
 \end{equation}
 where $\delta_{k,n}^{(R)}(z)$ and $\tilde\delta_{k,n}^{(R)}(z)$
 have the form (\ref{de^R}) and  due to (\ref{b_de^R}) satisfy the bound
 \begin{equation}\label{b-ti-de}
|\delta_{k,n}^{(R)}(z)|\le \frac{C}{n|\Im z|^2},\quad
|\tilde\delta_{k,n}^{(R)}(z)|\le\frac{Ck}{n|\Im z|^2}
+\frac{C\log^{1/2}n}{n^{1/2}|\Im z|^2}.
\end{equation}
 Thus, using the fact that
\[
\frac{2}{\pi}\int\frac{\mu^2\sqrt{4-\mu^2}}{(z-\mu)(\lambda-\mu)}
 d\mu=\frac{z^2\sqrt{z^2-4}}{z-\lambda}-(z^2+z\lambda+\lambda^2)+2,
\]
we get from (\ref{equR2})
\begin{equation}
R_{k,k}(z)=\left(z^2+a_k+
\delta_{k,n}^{(R)}(z)+\tilde\delta_{k,n}^{(R)}(z)+O(|\Im z|^{-1}n^{-1/2}\log n)\right)\frac{1}
{z^2\sqrt{z^2-4}}, \label{R3}
 \end{equation}
 where $R_{k,k}(z)$ is defined in (\ref{R_kk}) and we denote
\begin{equation}\label{a}
 a_k=\int \lambda^2[\psi^{(n)}_{k}(\lambda)]^2d\lambda-2.
\end{equation}
 Let us assume that $a_k>Cn^{-1/2}\log^{1/2} n$ with $C$ big enough. Then,
 using the bound (\ref{b-ti-de}) and the Rouchet  theorem, we get that
 $R_{k,k}(z)$ has a root in the circle of radius $\frac{1}{2}a_k^{1/2}$
 centered in the point $ia_k^{1/2}$. But,
 by definition (\ref{R_kk}),
\begin{equation}\label{Nev}
\Im R_{k,k}(z)\Im z<0,
\end{equation}
 so $R_{k,k}(z)$ cannot have zeros, when $\Im z\not=0$ and therefore
 we get $a_k\le Cn^{-1/4}\log^{1/2} n$. Similarly, if we assume that
 $a_k\le -Cn^{-1/4}\log^{1/2} n$
 we get that $\Im R_{k,k}(\frac{1}{2}|a_k|^{1/2}e^{i\pi/6})>0$, which also contradict to
 (\ref{Nev}).
Thus, we obtain that
\begin{equation}\label{b_a}
  |a_k|\le Cn^{-1/4}\log^{1/2} n.
\end{equation}
From (\ref{a}) and (\ref{R3}) we find
\begin{equation}\begin{array}{rcl}
(J^{(n)}_{k})^2+(J^{(n)}_{ k-1})^2&
=&\intd_{\sigma_e}\lambda^2(\psi^{(n)}_k(\lambda))^2d\lambda=2+a_k\\
((J^{(n)}_{k})^2+(J^{(n)}_{k-1})^2)^2+ (J^{(n)}_{k})^2(J^{(n)}_{k+1})^2&&\\
+(J^{(n)}_{k-1})^2(J^{(n)}_{k-2})^2&=&\intd_{\sigma_e}\lambda^4(\psi^{(n)}_k(\lambda))^2d\lambda=
{\fracd{1}{2\pi i}}\ds\oint_{L}\zeta ^{4}R_{k,k}(\zeta )d\zeta \\
&=&6+2a_k+O(n^{-1/2}\log  n).
\end{array}\label{sys1}\end{equation}
Using here the first equation for $k:=k\pm 1$ to express $(J^{(n)}_{k\pm 1})^2$
and $(J^{(n)}_{k-2})^2$ through
$(J^{(n)}_{k})^2$, we
obtain
\begin{equation}
J_{k}^2=1+\fracd{a_k}{2}+\fracd{a_{k+1}-a_{k-1}}{4}
\pm\bigg[\fracd{a_{k+1}+2a_k+a_{k-1}}{2}+O(n^{-1/2}\log n)\bigg]^{1/2}.
\label{sol1}\end{equation}
Combining this relation with (\ref{b_a}), we get the first statement of Lemma \ref{lem:1}.
The second statement follows from the first one and the first equation of (\ref{sys1}).

 $\square$

{\it Proof of Lemma \ref{lem:2}}

Choose $M=cn^{1/3}$, where the constant $c$ is small enough to provide
the condition
\begin{equation}\label{cond-M}
dC_2c<C_1/7,
\end{equation}
where $C_1$ and $C_2$ are the  constants from (\ref{exp_b}) and $d=\pi(2+\e)^{-1}$.
 This condition and (\ref{exp_b}) guarantee that for any $l,l':|l-l'|>n^{1/3}/6$ and any
$j:|j|<M$, $|t|\le 1$
\begin{equation}\label{exp_b1}
 |(e^{itdj\J^{(0)}})_{l,l'}|\le Ce^{dC_2M-C_1|l-l'|/4}\le Ce^{-C_1n^{1/3}/42}.
\end{equation}

Applying (\ref{d_exp}) three times we get (\ref{exp-v}) with
\[\begin{array}{rcl}
\P_{k-l}^{(\delta)}&=&c_1^{(\delta)}\delta_{k,l}+\intd_{s_1+s_2=1}ds_1ds_2
\sum_{j=\infty}^{\infty} v_j(ijd)
\left(e^{ijds_1\J^{(0)}}E^{(l)}
e^{ijds_2\J^{(0)}}\right)_{k,k+\delta},\\
\P^{(2,k,\delta)}_{l_1,l_2}&=&\intd_{\sum s_i=1}
ds_1ds_2 ds_3\sum_{j=-M}^M v_j(ijd)^2
\left(e^{ijds_1\J^{(0)}}E^{(l_1)}e^{ijds_2\J^{(0)}}E^{(l_2)}
e^{ijds_3\J^{(0)}}\right)_{k,k+\delta},\end{array}\]
\begin{equation}\label{rep-P}\begin{array}{rcl}
\P^{(3,k,\delta)}_{l_1,l_2,l_3}&=&\intd_{\sum s_i=1}
ds_1\dots ds_4\sum_{j=-M}^M v_j(ijd)^3\\&&
\left(e^{ijds_1\J^{(0)}}E^{(l_1)}e^{ijds_2\J^{(0)}}E^{(l_2)}
e^{ijds_3\J^{(0)}}E^{(l_3)}
e^{ijds_4\J^{(0)}}\right)_{k,k+\delta},\hskip 2cm\end{array}\end{equation}
\[\begin{array}{rcl}
r_k^{(\delta)}&=&\sumd_{l_1,\dots,l_4}\intd_{\sum s_i=1}
ds_1\dots ds_5
\sum_{j=-M}^Mv_j(ijd)^4\\&&\left(e^{ijds_1\J^{(0)}}\tilde\J
e^{ijds_2\J^{(0)}}\tilde\J
e^{ijds_3\J^{(0)}}\tilde\J e^{ijds_4\J^{(0)}}\tilde\J
e^{ijds_5(\J^{(0)}+\tilde\J)}\right)_{k,k+\delta}
\\&&+\intd_{s_1+s_2=1}ds_1ds_2
\sum_{|j|>M} v_j(ijd)
\left(e^{ijds_1\J^{(0)}}\tilde J
(e^{ijds_2\J^{(0)}}-e^{ijds_2(\J^{(0)}+\tilde\J)})\right)_{k,k+\delta},
\end{array}\]
where we denote by $E^{(l)}$ a matrix with entries:
\[E^{(l)}_{k,m}=\delta_{k,l}\delta_{m,l+1}+\delta_{k,l+1}\delta_{m,l}.\]
Using the Schwartz inequality, we have
\[\sum_{j}|j|^4|v_j|\le \bigg(\sum_{j}|j|^{10}|v_j|^2\bigg)^{1/2}
\bigg(\sum_{j\not=0}|j|^{-2}\bigg)^{1/2}\le C,
\]
Hence, using again the Schwartz inequality, we obtain
\begin{multline}\label{b_r_k}
|r_k^{(\delta)}|\le m^4_kd^4\sum_{|j|<M}|j|^4|v_j|+m_k
d\sum_{|j|>M}|j||v_j| \le C m^4_k+C m_kM^{-7/2}\le Cm^{4}_k,
\end{multline}
where the last inequality is valid because of the choice of $M$
and (\ref{m_N}).

To obtain (\ref{b-P}) we use the representation (see \cite{AS}):
\begin{equation}\label{exp(J_0)}
(e^{ijds\J^{(0)}})_{k,l}=\frac{1}{2\pi}\int_{-\pi}^{\pi}e^{ijds\cos
x} e^{i(k-l)x}dx =\mathbf{ J}_{k-l}(jds),
\end{equation}
where $\mathbf{ J}_{k}(s)$ is the Bessel function.
But it is well known (see, e.g. \cite{AS}) that
 the  Bessel functions satisfy the following recurrent relations:
\[
k\mathbf{J}_{k}(s)=\frac{s}{2}\bigg(\mathbf{ J}_{k+1}(s)+\mathbf{ J}_{k-1}(s)\bigg).
\]
Thus, e.g., the first sum in (\ref{b-P}) can be expressed via the
terms
\begin{multline*}
\sum_{|j|<M}v_j(ijd)^2\intd_{\sum s_i=1}
ds_1ds_2ds_3(djs_1+\alpha_1)(djs_1+\alpha_2)\\
\cdot\sum\nolimits'\mathbf{ J}_{k-l_1+\alpha_3}(djs_1) \tilde
x_{l_1}\mathbf{ J}_{l_1-l_2+\alpha_4}(djs_2)\tilde y_{l_2}
\mathbf{ J}_{l_2-k+\alpha_5}(2js_3),
\end{multline*}
where $\alpha_1,\dots\alpha_5$ can take the values $0,\pm1,\pm(\delta+1)$.
It is easy to see that any of these sums can be written in the form:
\[
(e^{ijds_2\J^{(0)}}X^{(\alpha_3)}e^{ijds_2\J^{(0)}}
Y^{(\alpha_4)}e^{ijds_2\J^{(0)}})_{k,k+\alpha_5},
\quad
X^{(\alpha)}_{k,l}=\delta_{k,\alpha+l}\tilde x_k,\quad
Y^{(\alpha)}_{k,l}=\delta_{k,\alpha+l}\tilde y_k,
\]
where evidently
\[
||X^{(\alpha)}||\le\max|\tilde x_k|,\quad
||Y^{(\alpha)}||\le\max|\tilde y_k|.
\]
Hence, similarly to (\ref{b_r_k}) we obtain
\[
\bigg|\sum\nolimits'\P^{(2,k,\delta)}_{l_1,l_2}(l_1-k)^2
\tilde x_{l_1}\tilde y_{l_2}\bigg|\le C||\tilde x||\,||\tilde y||
\sum_{|j|<M}|j|^4|v_j|\le C||\tilde x||\,||\tilde y||.
\]
The other inequalities in (\ref{b-P}) can be proved similarly.

We are left to prove (\ref{P^1}). Due to  representations
(\ref{rep-P}) and (\ref{exp(J_0)}),  we derive that
$\P^{(\delta)}$ can be represented in the form
 (\ref{P_l}) with
\[
 F^{(\delta)}(x)=\sum \P_l^{(\delta)}e^{ilx}
\]
Using (\ref{rep-P}) and (\ref{exp(J_0)}) we get
\begin{equation}\label{F^1}\begin{array}{rcl}
F^{(1)}(x)&=&c_1^{(1)}+\sumd_{j}(ijd)v_j\int_0^1 ds_1\sum_l\fracd{1}{4\pi^2}
\int_{-\pi}^{\pi}\int_{-\pi}^{\pi}e^{il(-x_1+x_2+x)}(1+e^{-i(x_1+x_2)})
\\
&&\cdot\exp\{2ijd[s_1\cos x_1+(1-s_1)\cos
x_2]\}dx_1dx_2\\
&=&c_1^{(1)}+\fracd{1}{2\pi}\int_{-\pi}^{\pi} \fracd{v(2\cos x_1)-v(2\cos(x_1-x))}
{\cos x_1-\cos(x_1-x)}(1+\cos (2x_1-x))dx_1\\
&=&c_1^{(1)}+\fracd{1}{2\pi}\int_{-\pi}^{\pi} v(2\cos x_1)\bigg(\frac{1+\cos (2x_1-x)}
{\cos x_1-\cos(x_1-x)}+\frac{1+\cos (2x_1+x)}
{\cos x_1-\cos(x_1+x)}\bigg)dx_1\\
&=&P(2\cos (x/2))+P(-2\cos (x/2)).
\end{array}\end{equation}
Representation (\ref{F^0}) can be obtained similarly.
Lemma \ref{lem:2} is proven.

\medskip
{\it Proof of Proposition \ref{pro:3}.} Let us remark first  that
all limiting expression in the r.h.s. of (\ref{p3.1}) and
(\ref{p3.1a}) correspond to infinite sums over $j$ in the
definitions (\ref{rep-P}) and infinite sums with respect to all
$l_i$. The estimates for the  remainder terms, which appears
because of the restriction of summation in (\ref{rep-P}) over
$|j|<M$, were obtained already in the proof of Lemma \ref{lem:2}.
And the remainders, which appear because of the replacement of
infinite sums by sums over $|l_i|<N_k$, can be estimated by
$O(e^{-C_1n^{1/3}/12})$ due to (\ref{exp_b1}). Thus  we are left
to compute  infinite over $l_i$ sums for
$\P^{(2,k,\delta)}_{l_1,l_2}$ and
$\P^{(3,k,\delta)}_{l_1,l_2,l_3}$

The first relation in (\ref{p3.1}) follows immediately from
(\ref{P_l}) and (\ref{F^1}).
To obtain the others let us  consider an infinite Jacobi
matrix $\J^{(\pi)}$ with
$J^{(\pi)}_{k,k+1}=J^{(\pi)}_{k+1,k}=(-1)^k$
and define
\begin{multline}\label{p3.2}
V_k(a,b)=V'(a\J^{0}+b\J^{(\pi)})_{k,k+1}=
\fracd{a}{2\pi(a^2-b^2)}\int_{\sigma}
V'(\lambda)\,\hbox{sign}\lambda
\fracd{(\lambda^2-4b^2)^{1/2}}{(4a^2-\lambda^2)^{1/2}}d\lambda\\+
\fracd{(-1)^kb}{2\pi(a^2-b^2)}\int_{\sigma} V'(\lambda)\,\hbox{sign}\lambda
\fracd{(4a^2-\lambda^2)^{1/2}}{(\lambda^2-4b^2)^{1/2}}d\lambda.
\end{multline}
It is easy to see, e.g., that
\[
\sumd_{l_1,l_2}\P^{(2,k,1)}_{l_1,l_2}(-1)^{l_1+l_2}=\frac{1}{2}\frac{\partial^2}{\partial
b^2}V_k(a,b)\bigg|_{a=1,b=0}=\frac{1}{2\pi}\int_{-2}^2
\frac{\lambda V'(\lambda) d\lambda}{\sqrt{4-\lambda^2}}-\frac{1}{\pi}\int_{-2}^2
\frac{ V'(\lambda) d\lambda}{\lambda \sqrt{4-\lambda^2}}=1.
\]
Here we have used (\ref{int0}) for the first integral and (\ref{P})
 the second.
Similarly
\[
\sumd_{l_1,l_2}\P^{(2,k,1)}_{l_1,l_2}(-1)^{l_1}=\frac{1}{2}\frac{\partial^2}
{\partial a\partial b}V_k(a,b)\bigg|_{a=1,b=0}=(-1)^k.\]
To compute the sum for $\P^{(3,k,1)}_{l_1,l_2,l_3}$ let us
observe that
\[\sumd_{l_1,l_2,l_3}\P^{(3,k,1)}_{l_1,l_2,l_3}(-1)^{l_1+l_2+l_3}
=\frac{1}{6}\frac{\partial^3}
{\partial b^3}V_k(1,b)\bigg|_{b=0}=\frac{(-1)^k}{2}\frac{\partial^2}
{\partial b^2}\fracd{1}{(1-b^2)}I(b)\bigg|_{b=0},
\]
where
\begin{multline*}
I(b)=\frac{1}{2\pi}\int_{\sigma} (V'(\lambda)-\lambda V''(0))\,\hbox{sign}\lambda
\fracd{(4-\lambda^2)^{1/2}}{(\lambda^2-4b^2)^{1/2}}d\lambda\\=
\frac{1}{2\pi}\int_{\sigma} V'(\lambda)\,\hbox{sign}\lambda
\fracd{(4-\lambda^2)^{1/2}}{(\lambda^2-4b^2)^{1/2}}d\lambda-V''(0)(1-b^2).
\end{multline*}
Differentiating this expression, one can easily get the expression
of (\ref{p3.1}).

To prove the last relation in (\ref{p3.1}) we use the symmetry
arguments. Indeed, according to (\ref{rep-P}),
\begin{multline*}
h(k,l_1)=\sum_{l_2}(\P^{(2,k,1)}_{l_1,l_2}+\P^{(2,k,1)}_{l_2,l_1})(-1)^{l_1+l_2}=
\sum_{j}v_j(ijd)^2\int_{s_1+s_2+s_3=1} ds_1 ds_2 ds_3\\
\bigg(u_{s_1}(k-l_1)f_{s_2,s_3}(k-l_1)+ u_{s_1}(k-l_1-1)f_{s_2,s_3}(l_1-k_1-1)\\ +
u_{s_1}(k-l_1)f_{s_1,s_2}(k-l_1) +u_{s_1}(k-l_1+1)f_{s_1,s_2}(l_1-k_1+1)\bigg),
\end{multline*}
where
\[
u_{s_1}(k-l)=(e^{ijds_1\J^{(0)}})_{k-l},\quad
f_{s_2,s_3}(l-k)=(-1)^l(e^{ijds_2\J^{(0)}}\J^{(\pi)}
e^{ijds_3\J^{(0)}})_{l,k}.
\]
Since both $u_{s_1}(k-l)$ and $f_{s_2,s_3}(l-k)$ are even functions
with
respect to $(l-k)$, after integration with respect to
$s_1,s_2,s_3$ we get that
\[
h(k,l_1)=h(k-l_1)=h(l_1-k)\Rightarrow \sum_{l_1}h(k,l_1)(k-l_1)=0.
\]
To prove (\ref{p3.1a}) we define
similarly to (\ref{p3.2})
\[\begin{array}{rcl}
V_k^{(0)}(b)=v^{(0)}(\J^{(0)}+b\J^{(\pi)})_{k,k}&=&\fracd{1}{\pi}\int_{\sigma}
\frac{V'(\lambda)\,\,\hbox{sign}\lambda}{(4b^2-\lambda)^{1/2}X(\lambda)}d\lambda
\\
&=&\fracd{1}{\pi}\int_{\sigma}
\frac{(V'(\lambda)-\lambda V''(0))\,\,\hbox{sign}\lambda}
{(4b^2-\lambda)^{1/2}X(\lambda)}d\lambda+V''(0).
\end{array}\]
Then
\[
\sumd_{l_1,l_2}\P^{(2,k,0)}_{l_1,l_2}(-1)^{l_1+l_2}=\frac{1}{2}\frac{\partial^2}{\partial
b^2}V_k^{0}(1,b)\bigg|_{b=0}=
\frac{2}{\pi}\int_{-2}^2
\frac{(V'(\lambda)-\lambda V''(0))d\lambda}{\lambda^3X(\lambda)}=2P_0(0).
\]
{\it Proof of Lemma \ref{lem:4}}.
Relation (\ref{t1.11}) can be written as
\begin{equation}\label{l4.2a}
d^{(2)}_k=2x_k^3+\tilde r_k,\quad \tilde r_k=x_k
k(2P_0(0)n)^{-1}+r_k,\quad |\tilde r_k|\le C_*(\tilde m_k|k|/n+\tilde m_k^4),
\end{equation}
where $C_*$ is independent of $N,n$ and we always can choose
$C_*>1$. If $\tilde m_k<k^{-1}$ for all $k>n^{1/3}$, then
(\ref{l4.1}) is fulfilled. If $\tilde m_k>k^{-1}$ for some
$k>7n^{1/3}$, we can apply Proposition \ref{p:d^2} to
$\{x_j\}_{|j|\le M}$, with $M=k$, $M_1=[n^{1/3}/2]$, $\varepsilon^3=
C_*\left(\tilde m_{k+2M_1}(k+2M_1)/n+m_{k+2M_1}^4\right)$,
$\varepsilon_1=m_{k+2M_1}$, because $M_1>2/3\e^{-1}$. Then, since
\[2\e_1M_1^{-2}=8\tilde m_{k+2M_1}n^{-2/3}<C_*\tilde m_{k+2M_1}(k+2M_1)/n<\e^3,\]
we obtain by (\ref{p5.0}) that $8\varepsilon^3=
8C_*\left(\tilde m_{k+2M_1}(k+2M_1)/n+m_{k+2M_1}^4\right)\ge m_k^3$. Therefore
at least one of the following inequalities holds
\begin{equation}\label{l4.2}
8C_*\tilde m_{k+2M_1}(k+2M_1)/n\ge m_k^3/2\quad\vee \quad8C_*m_{k+2M_1}^4\ge m_k^3/2
\end{equation}
Since according to Lemma \ref{lem:1} $|m_{k+2M_1}|\le Cn^{-1/8}\log^{1/4}n$ the second
inequality yields
\begin{equation}\label{l4.3a}
\tilde m_{k+2M_1}> 2\tilde m_k
\end{equation}
If the second inequality in (\ref{l4.2}) is false, then the first one holds. Write it as
\begin{equation}\label{l4.2b}
 \tilde m_{k+2M_1}\ge (16C_{*})^{-1}\tilde m_k \left[\tilde
m^2_kn/k\right]\left[ k/(k+n^{1/3})\right].
\end{equation}
Assume that for some $k>7n^{1/3}$
\begin{equation}\label{l4.3}
\tilde m^2_kn/k\ge 32C_{*}.
\end{equation}
Then (\ref{l4.2b})implies  (\ref{l4.3a}) and
\[
\tilde m^2_{k+2M_1}n/(k+2M_1)\ge 4\left(\tilde
m^2_kn/k\right)\cdot \left(k/(k+2M_1)\right)> 32C_{*}.
\]
Hence, we can repeat this procedure $l$ times with $l=[\log n]$.
Then we  obtain the inequality
\[
\tilde  m_{k+[\log n]M_1}>2^{[\log n]}\tilde m_k,
\]
which contradicts to Lemma \ref{lem:1}. Thus, (\ref{l4.3}) is
false and we have proved (\ref{l4.1}).

To prove (\ref{as+}) take any $k_0>n^{1/3}$ denote $\tilde x_k= x_{k-2k_0}$ and,
taking into account (\ref{t1.11}), apply (\ref{p5.2})
with $M=k_0$. Then since $f_k>(k_0/2P_0(0)n)$ we obtain (\ref{as+}).

\medskip

 {\it Proof of Lemma \ref{lem:5}}.

 Substituting in (\ref{equg}) $\phi(\lambda)=(z-\lambda)^{-1}$
 we get easily the equation
\begin{equation}\label{l5.3}
  g_{n+k,n}^2(z)-\int\frac{V'(\lambda)}{z-\lambda}\rho_{n+k,n}(\lambda)d\lambda
  =\delta_{n+k,n}(z)
\end{equation}
with $\delta_{n+k,n}(z)$ of the form (cf. (\ref{delta}))
\begin{equation}\label{l5.2a}
\delta_{n+k,n}(z)=\fracd{(J^{(n)}_{n+k})^2}{n^2}\left((R^2)_{n+k,n+k}(z)(R^2)_{n+k-1,n+k-1}(z)-
(R^2)^2_{n+k,n+k-1}(z)\right),
\end{equation}
where
$(R^2)_{k,j}=(z-\J^{(n)})^{-2}_{k,j}$. Here we have used the Christoffel-Darboux formula
in the numerator of the first integral in (\ref{l5.2}).
 Let us transform
\begin{multline}\label{l5.4}
  \int\frac{V'(\lambda)}{z-\lambda}\rho_{n+k,n}(\lambda)d\lambda=g_{n+k,n}(z)
  \left(zV''(0)+
  z^3\frac{V^{(4)}(0)}{6}\right)
  -\int\frac{V'(\lambda)}{\lambda}\rho_{n+k,n}(\lambda)d\lambda\\
  -z^2\int\frac{V'(\lambda)-\lambda V''(0)}{\lambda^3}\rho_{n+k,n}(\lambda)d\lambda
  -z^4\int\frac{V'(\lambda)-\lambda V''(0)
  -\frac{1}{6}\lambda^3V^{(4)}(0)}{\lambda^4(z-\lambda)}\rho_{n+k,n}(\lambda)d\lambda\\
  =g_{n+k,n}(z)  \left(zV''(0)+
  z^3\frac{V^{(4)}(0)}{6}\right)-c_{k,n}^{(0)}-z^2c_{k,n}^{(2)}-z^4c_{k,n}^{(4)}(z).
\end{multline}
Denote
\[
Q(\lambda)=\int\frac{V'(\lambda)-V'(\mu)}{\lambda-\mu}\rho(\mu)d\mu.
\]
 Taking the limit $n\to\infty$ in (\ref{l5.3}) and using (\ref{rho}), we get
 for any $\lambda\in[-2,2]$
\begin{equation}\label{Q}
\lambda^4P_0^2(\lambda)(\lambda^2-4)=[V'(\lambda)]^2-4Q(\lambda)\Rightarrow
Q(\lambda)=\frac{1}{4}\left([V'(\lambda)]^2
+\lambda^4P_0^2(\lambda)(4-\lambda^2)\right).
\end{equation}
Therefore, denoting $v^{(0)}(\lambda)=V'(\lambda)\lambda^{-1}$, we get
\begin{equation}\begin{array}{l}\label{l5.5}
c_{k,n}^{(0)}= Q(0)+\tilde c_{n}^{(0)}+c_k=\tilde
c_{n}^{(0)}+c_k,\\ \tilde c_{n}^{(0)}=\intd
v^{(0)}(\lambda)(\rho_{n,n}(\lambda)-\rho(\lambda))d\lambda, \quad
c_k=\pm n^{-1}\sum_{j=1}^{|k|} v^{(0)}(\J^{(n)})_{n\pm j,n\pm j}.
\end{array}\end{equation}
Here and below in the proof of Lemma \ref{lem:5} the sign $\pm$
corresponds to the sign of $k$. Repeating the argument of Lemma
\ref{lem:2} for the function $v^{(0)}(\lambda)$, we obtain
\[
v^{(0)}(\J^{(n)})_{n\pm j,n\pm j}=
x_j^2\sum_{|l_1|,|l_2|<|k|+n^{1/3}}{\cal
P}^{(2,k,0)}_{l_1,l_2}(-1)^{l_1+l_2}+
y_j\sum_{|l_1|<|k|+n^{1/3}}{\cal
P}^{(0)}_{l_1}+n^{1/6}O\left(\left(k/n\right)^{3/2}\right).
 \]
 Hence, using (\ref{p3.1a}),we get (\ref{c_k}).
 Now let us observe that
\[
\int\frac{V'(\lambda)-\lambda
V''(0)}{\lambda^3}\rho(\lambda)d\lambda=\frac{1}{2}
\frac{d^2}{d\mu^2}Q(\mu)\bigg|_{\mu=0} =\frac{1}{4}(V''(0))^2.
\]
Hence,
\[
c_{k,n}^{(2)}=\frac{1}{4}(V''(0))^2+\tilde c^{(2)}_n
\pm\sum_{j=1}^{|k|}
v^{(2)}(\J^{(n)})_{n\pm j,n\pm j},\quad\left(\,v^{(2)}(\lambda)=(V'(\lambda)-\lambda
V''(0))\lambda^{-3}\,\right),
\]
 where
\[
\tilde c^{(2)}_n=\int v^{(2)}(\lambda)(\rho_{n,n}(\lambda)-\rho(\lambda))d\lambda.
\]
Using Corollary \ref{cor:3} from Lemma \ref{lem:4}, we get
\[
v^{(2)}(\J^{(n)})_{n+k,n+k}=\int\frac{v^{(2)}(\lambda)d\lambda}{\sqrt{4-\lambda^2}}
+O\left(|k|/n\right)+O(n^{-2/3})=P_0(0)
+O\left(|k|/n\right)+O(n^{-2/3}).
\]
Therefore
\begin{equation}\label{l5.6}
c_{k,n}^{(2)}=\frac{1}{4}(V''(0))^2+P_0(0)\frac{k}{n}
+O\left(k^2/n^2\right)+ \tilde c^{(2)}_n.
\end{equation}
Now we apply (\ref{univ.1}) to
\[v^{(4)}(\lambda,z)=\frac{V'(\lambda)-\lambda V''(0)
  -\frac{1}{6}\lambda^3V^{(4)}(0)}{\lambda^{4}(z-\lambda)} .
  \]
We get
\begin{multline}\label{l5.7}
c_{k,n}^{(4)}(z)=\int v^{(4)}(\lambda,z)\rho(\lambda)d\lambda \pm
n^{-1} \sum_{j=1}^{|k|} v^{(4)}(\J^{(n)})_{n\pm j,n\pm j}+
 O\left(n^{-1/2}\log^{1/2}n/|\Im z|\right)\\
  =\frac{Q^{(4)}(0)}{4!}+ O(z)+O\left(n^{-1/2}\log^{1/2}n/|\Im z|\right)
 +O\left(k/n\right)\\
  =\frac{1}{4!}\left(24P_0^2(0)+2V'(0)V''(0)\right)+ O(z)
  +O\left(n^{-1/2}\log^{1/2}n/|\Im z|\right)+O\left(k/n\right),
\end{multline}
where the last equality follows from (\ref{Q}). Now, substituting
(\ref{l5.5})-(\ref{l5.7}) in (\ref{l5.4}), we find
\begin{multline}\label{l5.7b}
g_{n+k,n}(z)=\frac{1}{2}  \left(zV''(0)+
  z^3\frac{V^{(4)}(0)}{6}\right)-\left(-P_0^2(0)z^4-\frac{k}{n}P_0(0)z^2-c_k
 + \delta_{n+k,n}(z)-\tilde\delta_{n+k,n}(z)\right)^{1/2}
\end{multline}
with $c_k$ defined by (\ref{c_k}), $\delta_{n+k,n}(z)$ defined by
(\ref{l5.2a}) and
\begin{equation}\label{l5.7a}
\tilde\delta_{n+k,n}(z)=\tilde c_n^{(0)}+z^2\tilde c_n^{(2)}
+z^2\left(O(n^{-4/3})+O(k^2/n^2)\right) +O(z^5)+\frac{z^4}{|\Im
z|}O(n^{-1/2}\log^{1/2}n).
\end{equation}
Since (\ref{l5.1}) follows from (\ref{l5.7b}),  we are left to
estimate $\tilde c^{(0)}_n$, $\tilde c^{(2)}_n$ and
$\delta_{n+k,n}(z)$.

Taking into account (\ref{l5.2a}), to estimate $\delta_{n+k,n}(z)$
we need to estimate $(R^2)_{n+k,n+k}$ and
$(R^2)_{n+k,n+k-1}$.
Let us take $N'=k+\log^2n n^{1/3}$, and consider $\tilde\J(N')$ defined
by (\ref{ti-J}) and
\[
R^{(1)}(z)=(z-\tilde\J^{(0)}-\tilde\J(N'))^{-1}.
\]
Then,  using the resolvent identity (\ref{res-id}) and (\ref{exp_b_R}), we get
  for any $z:\Im z>n^{-1/3}$
\begin{equation}\label{l5.8a}
|(R^2)_{n+k,n+k}-(R^{(1)}*R^{(1)})_{k,k}(z)|\le Ce^{-C\log^2 n}.
\end{equation}
Applying the resolvent identity (\ref{res-id}) to $R^{(0)}=(z-\J^{(0)})^{-1}$
and $R^{(1)}(z)$ defined above, we get
\begin{multline}\label{l5.8}
|(R^{(1)}*R^{(1)})_{k,k}(z)-(R^{(0)}*R^{(0)})_{k,k}(z)|\\
\le
2\left((R^{(0)}(z)*R^{(0)}(\overline z))_{k,k}+ |\Im
z|^{-2}(R^{(0)}(z)*\ti J^2(N')*R^{(0)}(\overline z))_{k,k}
\right)\\ \le\frac{C}{|\Im z|}+\frac{C|k|}{n|\Im
z|^3}+\frac{2}{n|\Im z|^2}\sum
|R^{(0)}(z)_{k,j}|^2|k-j|\le\frac{C}{|\Im
z|}\left(1+\frac{|k|}{n|\Im z|^2}+ \frac{1}{n|\Im z|^3}\right).
\end{multline}
Here we have used  that for $|\Im z|\le 1$
\[
(R^{(0)}(z)*R^{(0)}(\overline z))_{k,k}= \sum|R^{(0)}_{k,j}(z)|^2=
\frac{1}{\pi}\int_{-2}^{2}\frac{d\lambda}{|z-\lambda|^2\sqrt{4-\lambda^2}}\le
\frac{C}{|\Im z|}.
\]
Substituting  (\ref{l5.8}) in (\ref{l5.2a}), we get the first
estimate in (\ref{l5.2}).

To estimate $\tilde c^{(0)}_n$ and $\tilde c^{(2)}_n$ we  subtract
from (\ref{l5.3})
 the same equation for $k:=k-1$ and multiply the result by $n$
 (see the proof of Lemma \ref{lem:1} for the details). Then we get
\begin{equation}\label{l5.9}
 2g_{n+k,n}(z)R_{n+k,n+k}(z)=\int\frac{V'(\lambda)}{z-\lambda}
 (\psi_{n+k}^{(n)}(\lambda))^2d\lambda
-\delta^{(R)}_{n+k,n}(z),
\end{equation}
where
\[
\delta^{(R)}_{n+k,n}(z)=\frac{1}{n}\sum_{j=1}^{\infty}R_{n+k,j}(z)R_{j,n+k}(z)-
\frac{2}{n}\sum_{j=1}^{n+k}R_{n+k,j}(z)R_{j,n+k}(z).
\]
Using the same trick as  in (\ref{l5.8a}), we get
\[
\delta^{(R)}_{k,n}(z)=
\frac{1}{n}\sum_{j=1}^{\infty}R^{(1)}_{k,j}(z)R^{(1)}_{j,k}(z)-
\frac{2}{n}\sum_{j=1}^{k}R^{(1)}_{k,j}(z)R^{(1)}_{j,k}(z)+O(e^{-C\log^2
n}).
\]
Besides, since  $R^{(0)}_{k,j}(z)$ is an even function of $(j-k)$,
we observe that
\[
0=\frac{1}{n}\sum_{j=-\infty}^{\infty}R^{(0)}_{k,j}(z)R^{(0)}_{j,k}(z)-
\frac{2}{n}\sum_{j=-\infty}^{k}R^{(0)}_{k,j}(z)R^{(0)}_{j,k}(z)+
\frac{1}{n}(R^{(0)}_{k,k}(z))^2,
\]
Hence, to estimate $\delta^{(R)}_{n+k,n}(z)$ it is enough to
estimate the difference between r.h.s. of the last two formulas.
Using (\ref{l5.8}) for the difference of the first sums, similar
bound for the difference of the second sums, and the bound
$|R^{(0)}_{k,k}(z)|=|z^2-4|^{-1/2}\le C$ for $|z|<1$, we get
\begin{equation}\label{l5.10}
|\delta^{(R)}_{n+k,n}(z)|\le\frac{C}{n|\Im
z|}\left(1+\frac{|k|}{n|\Im z|^2}+ \frac{1}{n|\Im z|^3}\right).
\end{equation}
Now performing transformations (\ref{l5.4}) for the integral in
the r.h.s. of (\ref{l5.9}), we can rewrite it as
\begin{equation}\label{l5.11}
R_{n+k,n+k}(z)(2g_{n+k,n}(z)-V''(0)z-V^{(4)}(0)z^3/6)=a_{k,n}^{(2)}z^2-a_{k,n}^{(0)}
+\delta^{(R)}_{n+k,n}(z)+O(z^4),
\end{equation}
where
\[\begin{array}{l}
a_{k,n}^{(0)}=v^{(0)}(\J^{(n)})_{n+k,n+k}=
2n^{-2/3}P_0(0)q_n^2(\frac{k}{n^{1/3}})+\frac{k}{2n}+O(n^{-1})\\
a_{k,n}^{(2)}=v^{(2)}(\J^{(n)})_{n+k,n+k}=
P_0(0)+O(n^{-2/3}).
\end{array}\]
Let us take $k>0$  and change the variable $z=\tilde\e\zeta$ with
$\tilde\e^2=k/P_0(0)n$ in (\ref{l5.11}). Then, using
(\ref{l5.7b}), we obtain from (\ref{l5.11})
\begin{equation}\label{l5.12}
R_{n+k,n+k}(\tilde\e\zeta)=\frac{\zeta^2+\frac{1}{2}
+2P_0(0)\frac{n^{1/3}}{k}q_n^2(\frac{k}{n^{1/3}})+
\tilde\e^{-2}P_0^{-1}(0)\delta^{(R)}_{n+k,n}(\tilde\e\zeta)+o(1)}
{2i\bigg(\zeta^4+\zeta^2+\tilde\e^{-4}P_0^{-2}(0)\left( c_k
+\tilde\delta_{n+k,n}(\tilde\e\zeta)-\delta_{n+k,n}(\tilde\e\zeta)\right)\bigg)^{1/2}}
:=\frac{R_1(\zeta)}{R_2^{1/2}(\zeta)}.
\end{equation}
In view of (\ref{l5.10})
\[
\tilde\e^{-2}|\delta^{(R)}_{k,n}(\tilde\e\zeta)|\to 0, \quad as\quad
\fracd{k}{n^{1/3}}\to \infty,\quad  \Im\zeta>d
\]
with any fixed $d$ (see (\ref{l5.10})). Besides, $q_n^2(x)\to 0$, as $x\to \infty$,
 because of (\ref{as+}).
Therefore there exists some fixed $l_0>0$, such that for $k>l_0n^{1/3}$
and any $\zeta:\Im\zeta>1/4$
\[
\bigg|2P_0(0)\frac{n^{1/3}}{k}q_n^2(\frac{k}{n^{1/3}})+
\tilde\e^{-2}P_1^{-1}(0)\delta^{(R)}_{n+k,n}(\tilde\e\zeta)\bigg|<\frac{1}{4}<
\min_{|\zeta-i/\sqrt 2|=1/4}\bigg|\zeta^2+\frac{1}{2}\bigg|.
\]
Then according to the Rouchet theorem $R_1(\zeta)$ has a root inside the circle
${\cal B}$ of radius $1/4$ centered at $i/\sqrt 2$.
Thus, if $R_2(\zeta)$ has no roots of the second order inside ${\cal B}$, then similarly to the
proof of Lemma \ref{lem:1} we obtain a contradiction with (\ref{Nev}). Therefore, using
the first inequality of (\ref{l5.2}), (\ref{l5.7a}) and (\ref{l5.10})
we conclude that there exists an absolute constant $C_0$, such that
\[
|\tilde c^{(0)}_n|,|\tilde c^{(2)}_n|^2\le C_0 n^{-4/3}.
\]
These bounds and (\ref{l5.7a}) prove the second estimate of
(\ref{l5.2}).

\section{Appendix}
{\it Proof of Proposition \ref{pro:2}}.
Using the spectral theorem and Proposition \ref{pro:1}, we get
\[ \bigg|v(\J^{(n)})_{n+k,n+k+\delta}-\tilde v(\J^{(n)})_{n+k,n+k+\delta}\bigg|=
\bigg|\int(v(\lambda)-\tilde
v(\lambda))\psi^{(n)}_{n+k}(\lambda)\psi^{(n)}_{n+k+\delta}(\lambda)d\lambda\bigg|\le Ce^{-nC\e}.
\]
Let us represent $\tilde v(\lambda)$ by its Fourier expansion
\begin{equation}\label{Four}
 \tilde v(\lambda)=\sum_{j=-\infty}^\infty v_je^{ijd\lambda},\quad
  d=\frac{\pi}{2+\e}.
\end{equation}
Then we have
\begin{equation}\label{Four1}
\tilde v(\J^{(n)})=\sum_{j=-\infty}^\infty v_je^{ijd\J^{(n)}}
 =\sum_{|j|\le cM} v_je^{ijd\J^{(n)}}+O(M^{-\ell+1/2}),
  \end{equation}
  where $c$ is some absolute constant which we will choose later.
The bound for the remainder term in the last formula follows from the
estimate
\[\bigg\|\sum_{|j|>cM} v_je^{ijd\J^{(n)}}\bigg\|\le \sum_{|j|>cM} |v_j|
\le
\bigg(\sum_{|j|>cM} |v_j|^2|j|^{2\ell}\bigg)^{1/2}
\bigg(\sum_{|j|>cM} |j|^{-2\ell}\bigg)^{1/2} \le
||v^{(\ell)}||_2(cM)^{-\ell+1/2}.
\]

Consider now $N'=[N+M]+1$ and denote by $\J^{(n,N')}$ the matrix  whose entries coincide with
that of $\J^{(n)}$ with the only exception $\J^{(n,N')}_{n\pm
N',n\pm N'+1}=0$. We will use the identity, valid for any matrices
$\J_1,\J_2$
\begin{equation}\label{d_exp}
e^{it\J_2}-e^{it\J_1}=\int_0^te^{i(t-s)\J_1}(\J_2-\J_1)e^{is\J_2}ds
\end{equation}
Let us take $|k|<N$ and apply (\ref{d_exp}) to
$\J^{(n)}_{n+k,n+k+\delta}$ and $\J^{(n,N')}_{n+k,n+k+\delta}$. Then we get
\begin{multline}\label{d_v}
v(\J^{(n)})_{n+k,n+k+\delta}-v(\J^{(n,N)})_{n+k,n+k+\delta}\\=
\int_0^tds\sum_{|j|\le M}v_j\sum_{\pm}\bigg((e^{ijd(t-s)\J^{(n,N')}})_{n+k,n\pm N'}
\J^{(n)}_{n\pm N',n\pm N'+1}(e^{is\J^{(n)}})_{n\pm N'+1,n+k+\delta}\\
+(e^{ijd(t-s)\J^{(n,N')}})_{n+k,n\pm N'+1}
\J^{(n)}_{n\pm N'+1,n\pm N'}(e^{is\J^{(n)}})_{n\pm
N',n+k+\delta}\bigg)+O(M^{-\ell+1/2}).
\end{multline}
Now we use the bound, valid for any
 Jacobi matrix $\J$ with coefficients $%
J_{k,k+1}=J_{k+1,k}=a_{k}\in \mathbf{R}$, $|a_{k}|\leq A$.
Then there exist positive constants $C_0,\,C_{1},C_2$, depending
on $A$ such that  the
matrix elements of  $e^{it\J}$ satisfy the inequalities:
\begin{equation}
|(e^{it\J})_{k,j}|\leq C_0e^{-C_1|k-j|+C_2t}.
\label{exp_b}
\end{equation}
This bound follows from the representation
\[(e^{it\J})_{k,j}=\frac{1}{2\pi i}\oint_le^{itz}\tilde R_{k,j}(z)dz,\]
where $\tilde R=(zI-\J)^{-1}$, and from (\ref{exp_b_R}).

Using (\ref{exp_b}) in (\ref{d_v}),  we get for any
$c<C_0(C_1d)^{-1}$
\[
 v(\J^{(n)})_{n+k,n+k+\delta}-v(\J^{(n,N')})_{n+k,n+k+\delta}=O((c M)^{-\ell+1/2}).
\]
Similarly (see definitions (\ref{J^0}) and (\ref{ti-J})):
\[
v(\J^{(0)}+\tilde \J)_{k,k+\delta}-v(\J^{(n,N')})_{n+k,n+k+\delta}=O((cM)^{-\ell+1/2}).
\]
These two bounds give us (\ref{v(J)}).

\medskip

\noindent {\it Proof of Proposition \ref{p:d^2}.} Assume that $ |\tilde x_k|>\varepsilon$
for some $k$.
 Without loss of generality we can assume that $\tilde x_k>0$. Then due
to (\ref{d^2>})
\[
\tilde x_{k+1}-2\tilde x_k+\tilde x_{k-1}>\tilde x_{k}^3.
\]
 Consider first the case when also $\tilde d^{(1)}_k=
\tilde x_{k+1}-\tilde x_k>0$. Then by induction for any $M-k>i>0$ we have
$\tilde d^{(1)}_{k+i}>\tilde d^{(1)}_{k}$, $\tilde x_{k+i}> \tilde x_{k}$ and
$\tilde d^{(2)}_{k+i}>\tilde x_{k}^3$. Hence
\[
\e_1>\tilde x_{k+M_1}>\tilde x_{k}+\tilde x_{k}^3M_1^2/2.
\]
 If $\tilde d^{(1)}_k<0$, then according to (\ref{d^2>})
 we have $\tilde x_{-k-1}>\tilde x_{k}$ and we obtain (\ref{p5.0})
 moving from $k$ in the negative direction.

Similarly,  assume that at some point $|k|\le M-2M_1$
\begin{equation}\label{p5.3a}
\tilde
d^{(1)}_k>4\max\{\varepsilon^2,(2\varepsilon_1M_1^{-2})^{2/3}\}=:4\mu^2.
\end{equation}
Since  $|\tilde d^{(2)}_k|\le 3\mu^3$ because of (\ref{d^2>}) and
the first inequality of (\ref{p5.0}),
 we have for any $|i|<i_0:=[2/(3\mu)]\le[2/(3\varepsilon)]\le M_1$
\[\tilde d^{(1)}_{k+i}>\tilde d^{(1)}_{k}/2\Rightarrow
\mu\ge |\tilde x_{k+si_0}|>i_0 \tilde
d^{(1)}_{k}/2=3\mu^{-1}d^{(1)}_{k},
\]
where $s=\hbox{sign}\tilde x_{k}$. The last inequality here
contradicts to (\ref{p5.3a}). Hence, (\ref{p5.3a}) is false and we
obtain the second inequality of (\ref{p5.0}).

To prove (\ref{p5.2}) observe that if we consider two $(2M+1)\times(2M+1)$
 Jacobi matrices $J^{(f)}$ and $J^{(d)}$ with entries
\[J^{(f)}=D^{(f)}-J^{(0)},\quad J^{(d)}=D^{(d)}-J^{(0)}
D^{(f)}_{k,j}=\delta_{k,j}(2+f_k),\quad D^{(d)}_{k,j}=\delta_{k,j}(2+d^2), \quad|k|\le 2M\]
then, using the Neumann expansion for their inverse we get
\begin{equation}\label{p5.3}
0\le (J^{(f)})^{-1}_{k,j}\le (J^{(f)})^{-1}_{k,j}\le(2d)^{-1}e^{-d|k-j|}.
\end{equation}
Hence, rewriting  (\ref{d^2}) as
\[
(J^{(f)}\tilde x)_k=-\tilde r_k+\delta_{k,2M}\tilde x_{2M+1}+\delta_{k,-2M}
\tilde x_{-2M-1}
\]
we get (\ref{p5.2}) from (\ref{p5.3}).

\medskip

{\it Proof of Proposition \ref{p:pain}}. It is evident that it is enough to
prove (\ref{stab}) for the case when  $2P_0(0)=1$ in (\ref{Pain}) and (\ref{stab}).
Hence, below we consider this case. For $x>0$ the statement is evident.
 Let $f(x)=\sqrt{-x/6}-q(x)$ and $x_0$ be the first negative root of $f$.
Since it is known (see \cite{HMcL}) that $q(x)=Ai(x)(1+o(1))$ as $x\to+\infty$,
we conclude that $q(0)>Ai(0)=0.355028...>3^{2/3}/6$ (for $Ai(0)$ see \cite{AS})
Besides, it is known that $q(x)>0$, $q'(x)<0$ (see \cite{HMcL}) and so
 \[\sqrt{-x_0/6}=q(x_0)>q(0)>3^{2/3}/6 \Rightarrow x_0<-3^{1/3}/2.\]
But for any point $x\le x_0<-3^{1/3}/2$ in which $q(x)\ge\sqrt{-x/6}$
\[q''(x)\le\sqrt{-x/6}(x-x/3)\le- \frac{2}{3}\sqrt{-x_0^3/6}<
-\left(4\sqrt{6(-x_0)^3}\right)^{-1}\le\left(\sqrt{-x/6}\right)''.\]
Therefore $f''(x)>0$ for $x\ge x_0$. Since by definition $f(x_0)=0$, $f'(x_0)\le 0$
(because $f(0)<0$ an $x_0$ is the first root of $f$) we  conclude that $f(x)<0$ for
any $x<x_0$ that contradicts to (\ref{as-Pain}).
 Thus we have proved that the left hand side of (\ref{stab}) is always
positive. But since it tends to infinity as $x\to\pm\infty$, we
conclude that there exists positive $\delta$, satisfying
(\ref{stab}).

\medskip

\noindent{\it Proof of Proposition \ref{pro:res}}.
 We use the estimate for
 matrix elements of the resolvent of an arbitrary Jacobi matrix $\mathcal{J}$,
 with entries $|\mathcal{J}_{j,j+1}|\le A$:
\begin{equation}
|\mathcal{ R}_{k,j}(z)|\leq {\frac{C_{1}'}{|\Im z|}}e^{-C_{2}'|\Im
z| |k-j|}, \label{C-T}
\end{equation}
 This estimate is similar to
well-known Combes- Thomas estimates for Schr\"{o}dinger operator
(see e.g.\cite{RS}).

Let $\mathcal{J}^{(k,M)}$ be the Jacobi matrix, whose entries
coincide with that for $\mathcal{J}$ with the only exceptions
$\mathcal{J}^{(k,M)}_{k\pm M,k\pm M+1}=\mathcal{J}^{(k,M)}_{k\pm
M+1,k\pm M}=0$, and
$\mathcal{R}^{(k,M)}=(z-\mathcal{J}^{(k,M)})^{-1}$. Then, by the
resolvent identity (\ref{res-id})
\[\mathcal{R}_{k,j}-\mathcal{R}^{(k,M)}_{k,j}=\sum_{\pm}
\left(\mathcal{R}^{(k,M)}_{k,k\pm M}\mathcal{J}_{k\pm M,k\pm
M+1}\mathcal{R}_{k\pm M+1,j}+ \mathcal{R}^{(k,M)}_{k,k\pm
M+1}\mathcal{J}_{k\pm M+1,k\pm M}\mathcal{R}_{k\pm M,j}\right).
\]
Since $\mathcal{J}^{(k,M)}$ has a block structure, its resolvent
$\mathcal{R}^{(k,M)}$ also has a block structure and its
coefficients $\mathcal{R}^{(k,M)}_{k,j}$ do not depend on
$\mathcal{J}_{j,j+1}$ with $|j-k|>M$. Hence, we can apply
(\ref{C-T}) to $\mathcal{R}^{(k,M)}_{k,j}$,
$\mathcal{R}^{(k,M)}_{k,k\pm M}$ and $\mathcal{R}^{(k,M)}_{k,k\pm
M+1}$. Then we get (\ref{exp_b_R}). Proposition \ref{pro:res} is
proved.

\medskip
 {\bf Acknowledgements.} The author is grateful to Prof. L.Pastur and
 Prof. A.Kuijalaars for the
 fruitful discussion.
 The author  acknowledges also the INTAS Research Network 03-51-6637 for the
 financial support.


\small

 \begin{thebibliography}{99}
\bibitem{AS} Abramowitz M., Stegun, I. \emph{Handbook of Mathematical
Functions}, Dover, N.Y., 1972

\bibitem{APS} S.Albeverio, L.Pastur and M.Shcherbina. On Asymptotic
Properties of the Jacobi Matrix Coefficients. {\it Matem. Fizika,
Analiz, Geometriya} {\bf 4}, 263-277 (1997)

\bibitem{APS2} Albeverio, S., Pastur, L., Shcherbina, M.: On the $1/n$
expansion for some unitary invariant ensembles of random matrices.
{\it Commun. Math. Phys.} {\bf 224},  271-305 (2001).

\bibitem{BI} Bleher, P., Its, A.: Double Scaling Limit in the Random Matrix
Model: the Riemann-Hilbert Approach. {\it Commun.Pure Appl.Math} \textbf{56},
433-516 (2003)

\bibitem{BPS} Boutet de Monvel, A., Pastur L., Shcherbina M.: On the
statistical mechanics approach in the random matrix theory.
Integrated density of states. {\it J. Stat. Phys. }\textbf{79},
585-611 (1995)

\bibitem{Bo-Br:91} M. Bowick, E. Brezin, Universal scaling of the tail of
the density of eigenvalues in random matrix models, {\it Phys.
Lett. }
\textbf{B 268%
}, 21-28 (1991).

\bibitem{CK} Claeys, T., Kuijalaars, A.B.J.: Universality of the double scaling limit
in random matrix models,
preprint arXiv:math-ph/0501074

\bibitem{De-Co:99} Deift, P., Kriecherbauer, T., McLaughlin, K., Venakides,
S., Zhou, X.: Uniform asymptotics for polynomials orthogonal with
respect to varying exponential weights and applications to
universality questions in random matrix theory.{\it Commun. Pure
Appl. Math.} \textbf{52} ,1335-1425 (1999)

\bibitem{De-Co:99a} Deift, P., Kriecherbauer, T., McLaughlin, K., Venakides,
S., Zhou, X.: Strong asymptotics of orthogonal polynomials with
respect to exponential weights. {\it Commun. Pure Appl. Math.
}\textbf{52} 1491-1552 (1999)

\bibitem{Dy} Dyson, D.J.: A Class of Matrix Ensembles. {\it J.Math.Phys.},\textbf{13} 90-107. 1972

\bibitem{Fo:93} P. Forrester, The spectrum edges of random matrix ensembles,
{\it Nucl. Phys.} \textbf{B 402} (1993) 709-728.

\bibitem{HMcL} Hastings, S.P.,  McLeod, J.B.: A boundary value problem
assiciated with the second Painleve transcendent and the Korteweg de
Vries equation.{\it Arch. Rational Mech.Anal},
\textbf{73}, 31-52 (1980)

\bibitem{Jo:98} Johansson, K., On fluctuations of eigenvalues of random
Hermitian matrices. Duke Math. J. \textbf{91}, 151-204 (1998)

\bibitem{LS} Levitan,B.M.,  Sargsyan, I.S. Introduction
to Spectral Theory, Translation Math. Monographs,39, AMS, Providence, RI, 1976

\bibitem{Me:91} M.L.Mehta, M.L.: \emph{Random Matrices}. New York: Academic
Press, 1991

\bibitem{Mus} Muskhelishvili N.I. Singular Integral Equations.
P.Noordhoff.-  Groningen 1953.

\bibitem{PS:97} Pastur, L., Shcherbina, M.: Universality of the local
eigenvalue statistics for a class of unitary invariant random
matrix ensembles. \emph{J. Stat. Phys.} \textbf{86}, 109-147
(1997)
\bibitem{PS3} Pastur, L., Shcherbina, M.:On the edge universality of the local
eigenvalue statistics of matrix models.  Matematicheskaya
fizika, analiz, geometriya \textbf{10}, N3, 335-365 (2003)

\bibitem{RS}  Reed,M., Simon,B.:\emph{Methods of Modern Mathematical
Physics, Vol.IV}, Academic Press: New York, 1978

\bibitem{Tr-Wi:94a} C.A. Tracy, H. Widom, Level spacing distributions and
the Airy kernel, \emph{Comm. Math. Phys}. \textbf{159},  151-174 (1994)
\end{thebibliography}
\end{document}